\documentclass[final, journal]{IEEEtran}
\usepackage{cite}

\usepackage[printonlyused]{acronym} % Acronyms

\usepackage[cmex10]{amsmath} % cmex10 option prevents amsmath from using a Type 3 font for math within footnotes.
\usepackage{amsthm} % AMS:Theorems
\usepackage{amssymb} % AMS:Symbols
\usepackage{algorithmic}
\usepackage{array}
\usepackage{cancel} % Strikeout part of the equation (\cancel, \bcancel, \xcancel)

\usepackage[caption=false,font=footnotesize]{subfig}
\usepackage[pdftex]{graphicx}
\graphicspath{{./figs/}}
\DeclareGraphicsExtensions{.pdf,.jpeg,.png}

\usepackage{color}

\usepackage{balance} % Balanced two-column mode
\usepackage{bm} % Makes the given argument bold

\usepackage{threeparttable} % Table
\usepackage{multicol} % Table: multicolumn
\usepackage{multirow} % Table: multirow
\usepackage{rotating} % Table: rotate text

\usepackage{upgreek} % upgreek greek letters
\usepackage{url} % Verbatim with URL-sensitive line breaks
\usepackage{verbatim} % Preformatted text blocks

\usepackage{mathtools} % Commands: coloneqq eqqcolon
\usepackage{mathrsfs} % Provides a \mathscr command

\usepackage{xfrac} % Split-level fractions

\usepackage{fancyhdr}
\usepackage{amsmath}

\usepackage{etoolbox}

\newcommand{\herm}{^{\mbox{\scriptsize H}}}

\newcommand{\tran}{^{\mbox{\scriptsize T}}}

\newcommand{\comb}{^{\mbox{\scriptsize c}}}

\newcommand{\cherm}{^{\mbox{\scriptsize cH}}}

\newcommand{\oLoop}{^{{\scriptsize (0)}}}
\newcommand{\iLoop}{^{{\scriptsize (i)}}}

\newcommand{\xiLoop}{^{{\scriptsize (i-1)}}}

\newcommand{\xiLoopComp}{^{{\scriptsize {*}(i-1)}}}

\newcommand{\iherm}{^{\scriptsize(i){\mbox{\scriptsize H}}}}

\newcommand{\HypTst}[2]{\raise2.5ex\hbox{\scriptsize$#1$} \hspace{-1.3em}\displaystyle\gtreqless\hspace{-1.2em}\raise-1.1em\hbox{\scriptsize$#2$}}

\newcommand{\cref}[1]{C\ref{#1}}
\acrodef{3-G}{3-Generation}
\acrodef{3GPP}{3rd Generation Partnership Project}
\acrodef{a.k.a}{also-known-as}
\acrodef{AAS}{active antenna system}
\acrodef{ADMM}{alternating direction method of multipliers}
\acrodef{AoA}{angle-of-arrival}
\acrodef{AoD}{angle-of-departure}
\acrodef{AHB}{Abel hybrid bound}
\acrodef{A-TS}{alternating Taylor\'s series}
\acrodef{A-LASSO}{\emph{adaptive}-least absolute shrinkage and selection operator}
\acrodef{AWGN}{additive white Gaussian noise} 
\acrodef{AGC}{Automated Gain Control}
\acrodef{BER}{Bit Error Rate}
\acrodef{BB}{Bhattacharyya Bound}
\acrodef{BF}{beamformer}
\acrodef{BFGS}{Broyden-Fletcher-Goldfarb-Shanno}
\acrodef{BS}{base-station}
\acrodef{BBU}{baseband processing unit}
\acrodef{cdf}{cumulative distribution function}
\acrodef{CEP}{Circular Error Probability}
\acrodef{CoMP}{Coordinated Multi-Point}
\acrodef{CIR}{Channel Impulse Response}
\acrodef{C-RDM}{Cross-Ranging Direction Matrix}
\acrodef{C-NLS}{Constrained Non-linear Least Squares}
\acrodef{CRLB}{Cram\'{e}r-Rao Lower Bound}
\acrodef{CSIT}{Channel State Information at the Transmitter}
\acrodef{CSI}{Channel State Information}
\acrodef{CP}{Cyclic-Prefix}
\acrodef{CB}{Coordinated Beamforming}

\acrodef{DC}{Distance Contraction}
\acrodef{DCT}{Distance Contraction Theory}
\acrodef{DE}{Distance Error}
\acrodef{DFT}{Discrete Fourier Transform}
\acrodef{DL}{Downlink}
\acrodef{DoA}{Direction-of-Arrival}
\acrodef{DSSS}{Direct Sequence Spread Spectrum}
\acrodef{DoF}{degree-of-freedom}
\acrodef{DPS}{Dynamic Point Selection}

\acrodef{EDF}{Euclidean Distance Function}
\acrodef{EDM}{Euclidean Distance Matrix}
\acrodef{EFIM}{Equivalent Fisher Information Matrix}
\acrodef{EHF}{Extremely High Frequency}
\acrodef{EKT}{Euclidean Kernel Transformation}
\acrodef{EK}{Euclidean Kernel}
\acrodef{ERII}{Equivalent Ranging Information Intensity}
\acrodef{ETSI}{European Telecommunications Standards Institute}
\acrodef{FIM}{Fisher Information Matrix}
\acrodef{FCC}{Federal Communications Commission}
\acrodef{FDD}{Frequency-Division-Duplex}
\acrodef{FP}{Fractional Program}
\acrodef{GLE}{Geometric-constrained Location Estimation}
\acrodef{G-LS}{Global Least Squares}
\acrodef{GNSS}{Global Navigation Satellite System}
\acrodef{GP}{Gaussian Process}
\acrodef{GP-LVM}{Gaussian Process Latent Variable Model}
\acrodef{GPS}{Global Positioning System}
\acrodef{GDC}{Global Distance Continuation}
\acrodef{GDOP}{Geometric Dilution of Precision}
\acrodef{GTRS}{Generalized Trust Region Subproblems}
\acrodef{HCRB}{Hammersley-Chapmann-Robbins Bound}
\acrodef{HB}{Hybrid Bound}
\acrodef{HOSVD}{High-Order Singular-Value-Decomposition}
\acrodef{HPBW}{half power beamwidth}
\acrodef{iid}{independent identically distributed}
\acrodef{ICT}{Information Communication Technology}
\acrodef{I-EKT}{Inverse Euclidean Kernel Transformation}
\acrodef{IoT}{Internet-of-Things}
\acrodef{IIoT}{Industrial Internet-of-Things}
\acrodef{iff}{if and only if}
\acrodef{IFT}{Inverse Fourier Transform}
\acrodef{ITU}{International Telecommunication Union}
\acrodef{I/N}{Interference-to-Noise}
\acrodef{JT}{Joint Transmission}
\acrodef{JT-CoMP}{Joint Transmission Coordinated Multi-Point}

\acrodef{KKT}{Karush-Kuhn-Tucker}
\acrodef{KLT}{Karhunen-Lo\'{e}ve Transform}
\acrodef{LASSO}{least absolute shrinkage and selection operator}
\acrodef{LARSO}{Least Absolute Residual and Selection Operator}
\acrodef{LBSs}{Location Based Services}
\acrodef{LM}{Levenberg-Marquardt}
\acrodef{LQ}{Link-Quality}
\acrodef{LS}{Least Squares}
\acrodef{LLS}{Linearized Least Squares}
\acrodef{LT}{Location-Tracking}
\acrodef{LTE}{Long Term Evolution}
\acrodef{LoS}{line-of-sight}
\acrodef{LOESS}{Locally weighted Scatterplot Smoothing}
\acrodef{LP}{Linear Programming}
\acrodef{MAP}{Maximum A Posteriori}
\acrodef{MOO}{Multiple-Objective Optimization}
\acrodef{MOS}{Mean Opinion Score}
\acrodef{MIMO}{multiple-input-multiple-output}
\acrodef{MU-MIMO}{multi-user multiple-input-multiple-output}
\acrodef{MU-MISO}{multi-user multiple-input-single-output}
\acrodef{MMSE}{Minimum Mean Squared Error}
\acrodef{mmWave}{millimeter-wave}
\acrodef{MSE}{Mean Square Error}
\acrodef{ML}{Maximum Likelihood}
\acrodef{MS}{mobile-station}
\acrodef{MDS}{Multidimensional Scaling}
\acrodef{MSPE}{Mean-Squared-Position-Error}
\acrodef{MUSIC}{MUltiple SIgnal Classification}
\acrodef{MRT}{maximum ratio transmission}
\acrodef{MURM}{Minimum User-Rate Maximization}

\acrodef{N-DC}{Negative-Distance Contraction}
\acrodef{NLOS}{non-line-of-sight}
\acrodef{NLS}{Non-Linear-Least Squares}
\acrodef{NP-hard}{Non-deterministic Polynomial-time hard}
\acrodef{NPRM}{Notice of Proposed Rulemaking}
\acrodef{OFDM}{orthogonal frequency-division multiplexing}
\acrodef{OMP}{Orthogonal Matching Pursuit}
\acrodef{OTDoA}{observed-Time-Difference-of-Arrival}
\acrodef{psd}{positive semi-definite}
\acrodef{PSD}{power spectral density}
\acrodef{pdf}{probability density function}
\acrodef{PDP}{Power Delay Profile}
\acrodef{P-DC}{Positive-Distance Contraction}
\acrodef{PE}{Position Error}
\acrodef{PEB}{Position Error Bound}
\acrodef{PREB}{position-rotation error bound}
\acrodef{PSM}{Positive Semi-definite Matrix}
\acrodef{POCS}{Projection On Convex Sets}
\acrodef{PCA}{Principal Component Analysis}
\acrodef{PPCA}{Probabilistic Principal Component Analysis}
\acrodef{QoD}{Quality-of-Design}
\acrodef{QoL}{Quality-of-Location}
\acrodef{QoS}{Quality-of-Service}
\acrodef{QoE}{Quality-of-Experience}
\acrodef{QoI}{Quality-of-Information}
\acrodef{R95}{95$\%$ Radius}
\acrodef{RBF}{Radial Basis Function}
\acrodef{RDM}{Ranging Direction Matrix}
\acrodef{REB}{Rotation Error Bound}
\acrodef{RF}{radio frequency}
\acrodef{R-GDC}{Range-Global Distance Continuation}
\acrodef{R-LS}{Regularized Least-Squares}
\acrodef{RSS}{Received Signal Strength}
\acrodef{RSSI}{Received Signal Strength Index}
\acrodef{RMB}{Reuven-Messer Bound}
\acrodef{RMSE}{root-mean-squared-error}
\acrodef{RII}{Ranging Information Intensity}
\acrodef{RR}{Ridge-Regression}
\acrodef{RRU}{remote radio unit}

\acrodef{SCE}{Sparse Channel Estimator}
\acrodef{SCM}{spatial channel model}
\acrodef{SDP}{Semi Definite Programming}
\acrodef{SIMO}{Single-Input-Multiple-Output}
\acrodef{SISO}{Single-Input-Single-Output}
\acrodef{SLAM}{Simultaneous Localization and Mapping}
\acrodef{SMACOF}{Stress-of-a-MAjorizing-Complex-Objective-Function}
\acrodef{SQP}{Sequential Quadratic Programming}
\acrodef{SR-LS}{Squared-Range Least-Squares}
\acrodef{SR-GDC}{Square Range-Global Distance Continuation}
\acrodef{SER}{Symbol-Error-Rate}
\acrodef{SB}{Stochastic Bound}
\acrodef{SNR}{Signal-to-Noise Ratio}
\acrodef{SHF}{Super High Frequency}
\acrodef{SINR}{signal-to-interference-plus-noise ratio}
\acrodef{SD}{Steepest-Descent}
\acrodef{SA}{Simulated-Annealing}
\acrodef{SR}{Squared Range}
\acrodef{SCA}{Successive Convex Approximation}
\acrodef{SOCP}{second-order cone program}
\acrodef{TDoA}{Time-Difference-of-Arrival}
\acrodef{TS}{Taylor's Series}
\acrodef{ToF}{Time-of-Flight}
\acrodef{ToA}{Time-of-Arrival}
\acrodef{TDD}{Time-Division-Duplex}
\acrodef{TTI}{Transmission Time Interval}
\acrodef{ULA}{uniform linear array}
\acrodef{UMTS}{Universal Mobile Telecommunications System}
\acrodef{URA}{uniform rectangular array}
\acrodef{UWB}{Ultra-WideBand}
\acrodef{UHF}{Ultra High Frequency}
\acrodef{UE}{user equipment}
\acrodef{WLS}{Weighted Least Squares}
\acrodef{WC}{Weighted Centroid}
\acrodef{WSRM}{Weighted Sum-Rate Maximization}
\acrodef{ZC}{Zadoff-Chu}
\acrodef{ZF}{Zero-Forcing}

\usepackage{dblfloatfix}    % To enable figures at the bottom of page

\IEEEoverridecommandlockouts

\def\@IEEEfigurecaptionsepspace{\vskip\abovecaptionskip\relax}%

\usepackage{optidef}
\usepackage[ruled,linesnumbered]{algorithm2e}
\newcounter{mytempeqncnt}

\newcommand{\subparagraph}{} % Solve: undefined control sequence in titlesec

\usepackage{bbm}

\begin{document}
%
%%\title{Highly-reliable Latency-aware mmWave Communication via Multi-point Connectivity}
\title{Latency-Constrained Highly-Reliable mmWave Communication via Multi-point Connectivity}
\author{
\IEEEauthorblockN{Dileep~Kumar,~\IEEEmembership{Student~Member,~IEEE,} Satya Joshi,~\IEEEmembership{Member,~IEEE,} and~Antti~T\"{o}lli,~\IEEEmembership{Senior~Member,~IEEE}}
\thanks{This work was supported by the European Commission in the framework of the H2020-EUJ-02-2018 project under grant no.~815056 ({5G}-Enhance) and Academy of Finland under grants no.~318927~({6Genesis Flagship}). }
\thanks{This article was presented in part at the IEEE Global Commun. Conf., Taipei, Taiwan, Dec. 2020~\cite{Dileep_Globecom2020}.  \textit{(Corresponding~author: Dileep~Kumar)} 
}
\thanks{The authors are with Centre for Wireless Communications, University of Oulu, FIN-90014 Oulu, Finland. (e-mail: dileep.kumar@oulu.fi; satya.joshi@oulu.fi; antti.tolli@oulu.fi).   }
}
%%%% Satya Joshi is with Nokia, Finland. (e-mail: satya.joshi@oulu.fi).
%%%% Authors Dileep Kumar, Satya Joshi and Antti T\"olli
\maketitle

%% -------------------------------------- ABSTRACT -------------------------------------- %%
%
\begin{abstract}
The sensitivity of millimeter-wave (mmWave) radio channel to blockage is a fundamental challenge in achieving low-latency and ultra-reliable connectivity. In this paper, we explore the viability of using coordinated multi-point (CoMP) transmission for a delay bounded and reliable mmWave communication. We propose a novel blockage-aware algorithm for the sum-power minimization problem under the user-specific latency requirements in a dynamic mobile access network. We use the Lyapunov optimization framework, and provide a dynamic control algorithm, which efficiently transforms a time-average stochastic problem into a sequence of deterministic subproblems. A robust beamformer design is then proposed by exploiting the queue backlogs and channel information, that efficiently allocates the required radio and cooperation resources, and proactively leverages the multi-antenna spatial diversity according to the instantaneous needs of the users. Further, to adapt to the uncertainties of the mmWave channel, we consider a pessimistic estimate of the rates over link blockage combinations and an adaptive selection of the CoMP serving set from the available remote radio units (RRUs). 
%%%%The coupled and non-convex constraints are handled via the Successive Convex Approximation (SCA) framework and Fractional Program (FP) techniques. 
Moreover, after the relaxation of coupled and non-convex constraints via the Fractional Program (FP) techniques, a low-complexity closed-form iterative algorithm is provided by solving a system of Karush-Kuhn-Tucker (KKT) optimality conditions. %%%, while ensuring average latency requirements. 
The simulation results manifest that, in the presence of random blockages, the proposed methods outperform the baseline scenarios and provide power-efficient, high-reliable, and low-latency mmWave communication.
\end{abstract}

%% -------------------------------------- INTRODUCTION -------------------------------------- %%
%
\section{Introduction}

The \ac{mmWave} communication is one of the key enabling technology for $5$th-generation ($5$G) and beyond cellular systems, which facilitates throughput-intensive and low-latency applications, such as \ac{IIoT}, factory automation, augmented reality, and autonomous driving~\cite{Bennis2020}. However, the full exploitation of the large available bandwidth at \ac{mmWave} frequencies, is mainly challenged by the sensitivity of directional radio links to the blockage, i.e., due to relatively higher penetration and path-losses. These lead to rapid degradation (i.e., strong dips) in the received signal strength, and thus results in intermittent connectivity~\cite{MacCartney-RapidFading-2017, AndreevBlockage2018}. For example, a human blocker can obstruct the dominant links for hundred of milliseconds, and may lead to disconnecting the ongoing communication session, which severely impacts the network's reliability~\cite{MacCartney-RapidFading-2017, AndreevBlockage2018}. Moreover, adapting to unpredictable blockage demands critical latency and signalling overhead, i.e., searching for an unblocked direction to re-establish the communication link~\cite{Shariatmadari-Linkadaptation-2016}. Therefore, unless being addressed properly, the blockage appears as the main bottleneck that hinders in achieving low-latency and reliable \ac{mmWave} connectivity.

To tackle the \ac{mmWave} radio channel uncertainties, the use of macro-diversity via \ac{CoMP} has gained great interest. In particular, the \ac{JT}-\ac{CoMP} connectivity, where each \ac{UE} is coherently served by multiple spatially distributed \acp{RRU}~\cite{Dileep_TWC2021, Kaleva-DecentralizeJP-2018}. Further, \ac{CoMP} schemes are also considered in recent \ac{3GPP} releases~\cite{NR3GPP-MultiConnectivity}, and it is envisioned that the use of multi-antenna spatial redundancy via geographically separated transceivers will be of high importance in the \ac{mmWave} based deployment scenarios. As a proof-of-concept, Qualcomm has recently implemented a \ac{CoMP} testbed at \ac{mmWave} frequencies~\cite{qualcomm}, and empirically shown the coverage and capacity improvements via flexible deployments in  {IIoT} and factory automation scenarios.

\subsection{Prior Work}
\label{subsec:PriorWork}
The \ac{CoMP} transmission and reception are mainly employed to enhance the system throughput, generally for the cell-edge users due to adverse channel conditions (e.g., higher path-loss and interference from neighboring \acp{RRU}). The \ac{CoMP} techniques, such as, \ac{JT}, \ac{CB} and \ac{DPS} are standardized in \ac{3GPP}~\cite{NR3GPP-MultiConnectivity}, and has been widely studied in past decade under the context of legacy $4$G systems~\cite{Antti-OntheValue-2009, irmer-coordinated-2011, Nigam-CoMP}. For example, in~\cite{Nigam-CoMP}, it is shown that \ac{JT}-\ac{CoMP} increases the coverage by up to $24\%$ for cell-edge users and $17\%$ for general users  compared to non-cooperative~scenarios.
Recent studies have also considered the \ac{CoMP} schemes in the mmWave frequencies~\cite{MacCartney-BSDiversityJournal-2019, Maamari-CoverageinmmWave-2016, Andreev-Multiconnectivity-2019, Andreev-DegreeOfMulti-2019}.

In~\cite{MacCartney-BSDiversityJournal-2019}, from extensive real-time measurements, the authors showed a significant coverage improvement by simultaneously serving a user with spatially distributed transmitters. The network coverage gain for the \ac{mmWave} system with multi-point connectivity, in the presence of random blockages, was also confirmed in~\cite{Maamari-CoverageinmmWave-2016, Andreev-Multiconnectivity-2019} using stochastic geometry tools. The gains of macro-diversity for the achievable rate and outage probability were quantified in~\cite{Andreev-DegreeOfMulti-2019}, and it shown that \ac{CoMP} connectivity, at the minimum of four spatially distributed links, can provide up to $76\%$ capacity gains.
However, \ac{CoMP} techniques were still devised with the sole scope of enhancing the capacity and coverage~\cite{MacCartney-BSDiversityJournal-2019, Maamari-CoverageinmmWave-2016, Andreev-Multiconnectivity-2019, Andreev-DegreeOfMulti-2019}, that inadvertently ends up relegating the stringent latency and reliability requirements, which is industrial-grade critical IoT applications (e.g., the factory automation scenarios). 
A step towards this direction is introduced in~\cite{Dileep_TWC2021}, where we provide reliable transmission schemes in the presence of random blockages, by preemptively underestimating the achievable rates over the potential link blockage combinations.   
However, these algorithms are still designed for the static case, i.e., the resource allocation problem for a given instance is studied without taking into account network dynamics and stringent latency conditions. %%%%%under the perfectly available channel is studied. 
Hence, these algorithms are not always applicable for, e.g., long-term time-dependent dynamics network deployment scenarios, and for retaining  stable connectivity while satisfying the user-specific latency requirements. 
%%%%%under the uncertainties of \ac{mmWave} radio channel. 
%
%
%%%%%It is worth highlighting that a system can easily achieve a desired level of reliability by sequential data transmission, i.e., by transmitting the same data packages until a receiver acknowledges the correct receptions over a dedicated feedback link~\cite{Johansson-Radioaccess-2015}. However, in the presence of random link blockages, high penetration and path-losses, the \ac{mmWave} feedback links are inherently unreliable and, hence, require redundant retransmissions. On the other hand, allowable latency dictates a strict upper limit on the number of retransmission attempts~\cite{Shariatmadari-Linkadaptation-2016}. 
% 
Furthermore, in low-latency industrial automation applications, the cycle time is shorter than the fading channel coherence time, which rules out the viability of automatic repeat request (ARQ)-based techniques for, e.g.,  delay bounded critical IoT applications~\cite{Harish-ARQ, Zheng_IoTJ_2019}.

Thus, the limitations of retransmission events and the difficulty of accurate estimation of random blocking events motivate us to develop highly-reliable latency-constrained transmission strategies for dynamic networks. %%%, that can retain stable and resilient \ac{mmWave} connectivity while satisfying the user-specific latency requirements. %%%%under the uncertainties of \ac{mmWave} radio channel and random blockers. 
Specifically, we investigate on, how to use queue backlogs and channel information at the transmitter to efficiently allocate the required radio and cooperation resources, and to proactively exploit the multi-antenna spatial diversity according to the instantaneous needs of the users for a dynamic \ac{mmWave} access networks.

%%%%\emph{The core question this paper tries to answer is how to use queue backlogs and channel state information at the transmitter to efficiently allocate the required radio and cooperation resources, and to proactively exploit spatial diversity according to the instantaneous needs of the users.}

\subsection{Contributions}
\label{subsec:Contribution}
We develop a robust downlink transmission strategy, tailored for a \ac{JT}-\ac{CoMP} based dynamic networks, satisfying the user-specific latency requirement while retaining stable and resilient connectivity under the uncertainties of \ac{mmWave} radio channel.  %%% and random blockers. 
Specifically, we consider a time-average sum-power minimization problem subject to maximum allowable queue length constraint in the presence of random blockages. The long-term time-average stochastic problem is transformed into a sequence of deterministic and independent subproblems using the Lyapunov optimization~\cite{neely2010stochastic}. The coupled and non-convex constraints are approximated with the sequence of convex subsets by using the \ac{FP} techniques~\cite{FP_Shen_2018}. Further, the proposed \ac{FP} based relaxations allow an efficient implementation of closed-form iterative beamformer design that enables tailored complexity and processing performance.  %%%% depending on the specific implementations.    %%% that allow tailored performance depending on the specific implementation constraints..... %%% function to a stationary point solution.  
%%%\textbf{Add here brief pros and cons} %The underlying convex subproblem 
%%%% is then iteratively solved until the convergence of objective function to a stationary point solution, and that 
The main contributions of this paper are summarized as follow:  
\begin{itemize}
    \item A robust transmit beamformer design is proposed by utilizing the multi-antenna spatial diversity and geographically separated transceivers in \ac{CoMP} connectivity scenarios. The time-average sum-power is minimized 
    %%%%under the probabilistic queue length constraints, 
    while ensuring the latency requirements, 
    where, for each user, a pessimistic estimate of the rates overall possible subset combinations of potentially blocked links is considered~\cite{Dileep_TWC2021}. Thus, managing mutually coupled link blocked combinations is more challenging than conventional constrained optimization.
    %%%among serving \ac{CoMP} \acp{RRU} is considered. 
    %
    \item To adapt with the uncertainties of the \ac{mmWave} radio channel, a proactive and dynamic selection of the \ac{CoMP} serving set from the available \acp{RRU} is proposed, e.g., by exploiting the queue backlogs and channel  information. This preemptive rate estimate and dynamic modeling of the serving subset is shown to greatly improve the reliability and average sum-power performance while ensuring the user-specific latency requirement. 
    \item After the relaxation of coupled and non-convex constraints via the Fractional Program (FP) techniques, a low-complexity robust beamformer design framework with specific performance and convergence characteristics is proposed by solving a system of closed-form \ac{KKT} optimality conditions and parallel update of beamforming vectors corresponding to the spatially distributed \acp{RRU}.
    %%%for the sum-power minimization problem, while ensuring average latency requirements.  
    This leads to  practical and computationally efficient implementation for, e.g., hardware constrained IoT devices with limited processing capabilities in IIoT and  factory automation scenarios.
\end{itemize}

The paper is an extended version of our previously published conference paper~\cite{Dileep_Globecom2020}. It is worth highlighting, in~\cite{Dileep_TWC2021} we studied reliable downlink transmission schemes for \ac{WSRM} problem. However, the algorithms were designed without taking into account stringent user-specific latency requirements and network dynamics. Hence, these schemes are not applicable to latency constrained dynamic mobile access scenarios. Motivated by the above highlighted concerns, we explored the viability of using \ac{CoMP} transmission for a delay bounded and reliable \ac{mmWave} communication in~\cite{Dileep_Globecom2020}. Nevertheless, the beamformer designs in~\cite{Dileep_Globecom2020, Dileep_TWC2021} were still based on the well-known \ac{SCA} framework~\cite{beck2010sequential}. Further, the pessimistic estimate of the achievable rates in~\cite{Dileep_Globecom2020, Dileep_TWC2021} were also modeled over the fixed subset combinations of serving set of \acp{RRU}, and that may lead to inefficient utilization of network resources. 
Compared to our previous works~\cite{Dileep_Globecom2020, Dileep_TWC2021}, in this paper, we have included the following additional notable  contributions that provide more complete coverage and analysis in dynamic \ac{mmWave} access networks. Specifically, the core question this paper tries to answer is how to use queue backlogs and channel information at the transmitter to efficiently allocate the required radio and cooperation resources, and to proactively exploit multi-antenna spatial diversity according to the instantaneous needs of the users in dynamic mobile network conditions. Further, we extend the standard \ac{FP} quadratic transform  techniques~\cite{FP_Shen_2018}, i.e., to take into consideration JT-CoMP connectivity and provide a novel grouping of a multitude of potentially coupled and non-convex SINR conditions, that raise from the link blockage subset combinations. The proposed extension facilitates a low-complexity iterative algorithm, where each step is computed in closed-form expressions. All of the aforementioned results have been further improved and extended in this paper while accounting network dynamics and link blockages. 
We provide comprehensive and detailed numerical examples to evaluate the performance advantage of the proposed solutions. Specifically, we quantify the underlying trade-offs in terms of average sum-power, achievable rates, and reliable connectivity, under the uncertainties of \ac{mmWave}  channel and random blockages.

\subsection{Organization and Notations}
 \vspace*{-2pt}
\label{subsec:Organization-Notations}
The remainder of this paper is organized as follows. In Section~\ref{sec:model}, we illustrate the system, blockage, and network queuing model, as well as, provide the formulation of the problem. Section~\ref{sec:DynamicAlgo_Lyapunov} provides a dynamic control algorithm. In Section~\ref{subsec:SINR_Approximation}, we describe the proposed beamformer designs. Section~\ref{subsec:DynSubSet_Selection} provides an algorithm for dynamic serving set selection and theoretical analysis of outage. The validation of our proposed methods with the numerical results are presented in Section~\ref{sec:Sim-Result}, and conclusions are given in Section~\ref{sec:conclusion}. 

\noindent \textit{Notations}: In the following, we denote vectors and matrices with boldface lowercase and uppercase letters, respectively. The inverse, conjugate transpose and transpose operation is represented with the superscript $(\cdot)^{\scriptsize -1}$, $(\cdot)\herm$ and $(\cdot)\tran$, respectively. Cardinality of a set $\mathcal{X}$ is denoted with $|\mathcal{X}|$. The norm and the real part of a complex number is represented with $|\cdot|$ and $\Re\{\cdot\}$, respectively. Notation $\mathbb{C}^{M \! \times \! N}$ is a ${M \! \! \times \! \! N}$ matrix with elements in the complex field. $[\mathbf{a}]_n$ is the $n${th} element of~$\mathbf{a}$, and $(a)^{\scriptsize +}\!\triangleq\!\mathrm{max}(0,a)$. Notation $\nabla_{\mathbf{x}}y(\mathbf{x})$ represent the gradient of $y(\cdot)$ with respect to~$\mathbf{x}$.

%% -------------------------------------- SYSTEM MODEL -------------------------------------- %%
%

\begin{comment}
%% ----- %%%% ----- %%
\begin{figure}[t]
\setlength\abovecaptionskip{-0.2\baselineskip}
\centering
\includegraphics[trim=0.0cm 0.0cm 0.0cm 0.0cm, clip, width=0.95\linewidth]{figsPaper/FactoryScenario.PNG} 
\caption{The propagation environment between a RRU-user pair.} 
\label{fig:Factory-Model}
\end{figure}
%% ----- %%%% ----- %%

\end{comment}

 \vspace*{-2.5pt}
\section{System Model}
 \vspace*{-1.5pt}
\label{sec:model}

\begin{figure}[t]
\setlength\abovecaptionskip{-0.2\baselineskip}
\centering
\includegraphics[trim=0.0cm 0.0cm 0.0cm 0.0cm, clip, width=0.75\linewidth]{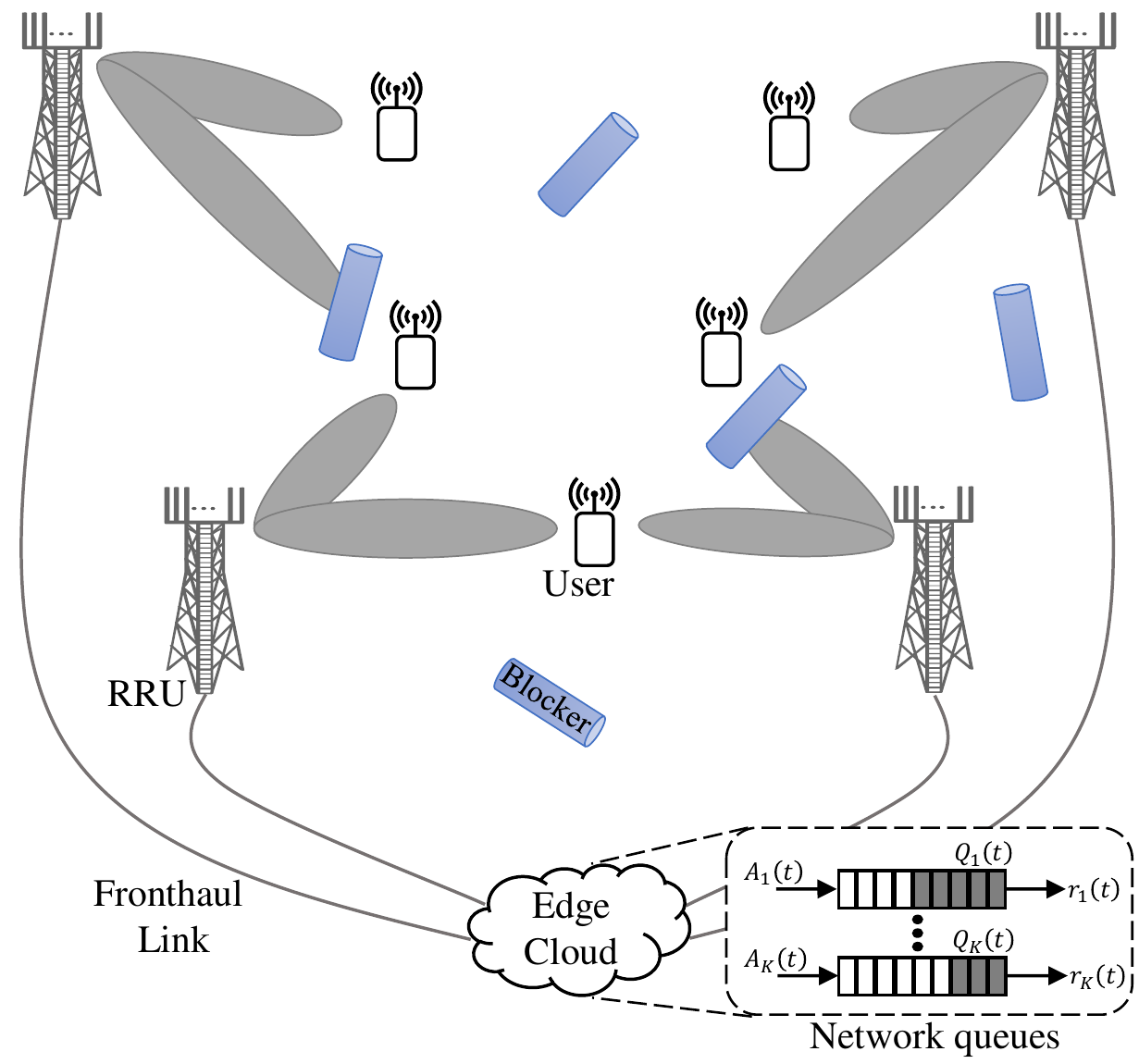} 
\caption{C-RAN with network queues, $B$~transmitters (RRUs) and $K$~receivers (UEs) in the presence of randomly distributed blockers.} 
\label{fig:System-Model}
 \vspace*{-12pt}
\end{figure}

We consider a downlink transmission in a \ac{mmWave} based cloud (or centralize) radio access network (C-RAN) architecture, where all \acp{RRU} are connected to the edge cloud by the fronthaul links, as illustrated in Fig.~\ref{fig:System-Model}. Each RRU is equipped with $N$~transmit antennas. We use $\mathcal{K}~\!=~\!\{ 1, 2,\ldots, K \}$ to denote the set of all single antenna \acp{UE} (also regarded as IoT devices), and $\mathcal{B}~\!=~\!\{ 1, 2,\ldots, B\}$ to denote the set of all \acp{RRU}. Further, the set of \acp{RRU} that serve $k$th \ac{UE} is denoted by $\mathcal{B}_k$, such that $\mathcal{B}_k~\!\subseteq~\!\mathcal{B}, \ \forall k\in\mathcal{K}$. We assume \ac{JT}-\ac{CoMP} connectivity~\cite{ Kaleva-DecentralizeJP-2018}, where each user~$k$ receives a  synchronous signal from its serving \acp{RRU}~$\mathcal{B}_k \ \forall k$. 
We assume that the network operates in a time-slotted manner, and slots are normalized to an integer value, e.g., $t\!\in\!\{1,2,\ldots\}$. Further, we assume that all \acp{RRU} use the same time-frequency resources for data transmission.

Let $\mathbf{f}_{b,k}(t) \in \mathbb{C}^{N \times 1}$ denote the transmit beamforming vector from $b$th~\ac{RRU} to $k$th~\ac{UE}. Then, the received signal
%%%%%\footnote{The methods can be extended to, e.g., hybrid analog-digital beamforming, where the channel between \ac{RRU}-\ac{UE} pair $\{\mathbf{h}_{b,k}\}$ can be considered as an effective channel obtained after the analog beamforming stage. } 
$y_k(t)$ at $k$th \ac{UE} during time slot~$t$ can be expressed as
\begin{align}
\label{eq:Rx-Signal}
y_k(t) = & {\textstyle \sum\limits_{b\in\mathcal{B}_k}} \mathbf{h}_{b,k}\herm(t) \mathbf{f}_{b,k}(t) d_{k}(t) \nonumber \\ 
& \qquad + \!\! {\textstyle \sum\limits_{u \in \mathcal{K} \backslash k} \sum\limits_{g\in\mathcal{B}_{u}}} \mathbf{h}_{g,k}\herm(t) \mathbf{f}_{g,u}(t) d_{u}(t) + {w}_k(t) ,
\end{align}
where $\mathbf{h}_{b,k}(t) \in \mathbb{C}^{N \times 1}$ is the channel vector between \ac{RRU}-\ac{UE} pair $(b,k)$. Notation $d_k(t)$ is data symbol associated with $k$th~\ac{UE}, and ${w}_k(t)\in\mathcal{CN}(0,\sigma_k^2)$ is circularly symmetric additive white Gaussian noise (AWGN). Moreover, we assume that data symbols are normalized and independent, i.e., $\mathbb{E}\{|d_k(t)|^2\}=1$ and $\mathbb{E}\{d_k(t) d_u^*(t)\}=0$ for all $k,u\in\mathcal{K}$.
The received \ac{SINR} of $k$th \ac{UE} during time slot~$t$ can be expressed as  
\begin{equation}
\label{eq:SINR}
\Gamma_k(\mathbf{F}(t)) = \frac{\Bigl| {\textstyle \sum\limits_{b\in \mathcal{B}_k}} \mathbf{h}_{b,k}\herm(t) \mathbf{f}_{b,k}(t) \Bigr|^2  }{\sigma_k^2 + {\textstyle \sum\limits_{u \in \mathcal{K} \backslash k}} \Bigl| {\textstyle \sum\limits_{g\in \mathcal{B}_{u}}} \mathbf{h}_{g,k}\herm(t) \mathbf{f}_{g,u}(t) \Bigr|^2 } ,
\end{equation}
where $\mathbf{F}(t) \triangleq [\mathbf{f}_{1,1}(t), \mathbf{f}_{1,2}(t), \dots, \mathbf{f}_{B,K}(t)]$.

%% ----- %%%% ----- %%
%% ----- %%%% ----- %%

%% ----- %%%% ----- %%%% ----- %%%% ----- %%
%
\subsection{Blockage Model and Achievable Rate}
\label{subsec:Blockage}
In \ac{mmWave} frequency band, the radio channel is spatially sparse due to low-scattering, reduced diffraction, and higher penetration and path losses~\cite{MacCartney-RapidFading-2017, AndreevBlockage2018}. Hence, a mmWave communication link is inherently unreliable due to its susceptibility to blockages. The channel measurements in a typical \ac{mmWave} outdoor scenarios have empirically shown that a link outage occurs with $20\% - 60\%$ probability~\cite{MacCartney-RapidFading-2017}, and may lead to a $10-$fold decrease in the achievable sum-rate performance~\cite{AndreevBlockage2018}. Therefore, unless being addressed properly, the random link blockage appears as the main bottleneck hindering the full exploitation of the \ac{mmWave} radio channel.
Thus, to characterize the aforementioned uncertainties of the \ac{mmWave} channel, we consider a probabilistic binary blockage model~\cite{Bai-Blockage-2014, Renzo-2015-BinaryChannel}. Specifically, the radio channel $\mathbf{h}_{b,k}(t)$ between \ac{RRU}-\ac{UE} pair $(b,k)$ can have one of two states, i.e., it is either fully-available or completely blocked. Furthermore, we consider link specific blockage, 
%%%%\footnote{This is reasonably accurate assumption of a \ac{mmWave} system, especially, when the blockers are not very large and reasonably closer to the \ac{UE}~\cite{Renzo-2015-BinaryChannel}.}
and the blocking of channel $\{\mathbf{h}_{b,k}=\mathbf{0}\}_{b\in\mathcal{B}, k\in\mathcal{K}}$ for all $t$ is independent. {The methodology can easily be extended to more elaborate channel blocking models, e.g., by considering the distance and the spatio-temporal correlations~\cite{Hriba-CorrelatedBlocking-2019}. In fact, this is an interesting topic for future extensions.} %%%%%Similar blockage models have been extensively studied, e.g., in~\cite{Dileep_Globecom2020, Dileep_TWC2021, Renzo-2015-BinaryChannel}. 

Similarly to~\cite{Dileep_TWC2021}, for improving system reliability and avoid outage under the uncertainties of mmWave channel, we preemptively underestimate the achievable \ac{SINR} assuming that a portion of \ac{CoMP} links would be blocked during the downlink data transmission phase. This is specifically required in the \ac{mmWave} communication because of dynamic blockages, which are, in general, not possible to track during the channel estimation phase.
Let \ac{BBU} assume that each user~$k$ have at least $L_k(t)\in\bigl[1, \ |\mathcal{B}_k|\bigr]$ available links (i.e., unblocked \acp{RRU}). Then, we allow BBU to proactively model the lower-bound of achievable \ac{SINR} over all possible subset combinations, e.g., by excluding the potentially blocked links, and allocate the pessimistic rate to users.
%%%%%%, and allocate the rate to users such that transmission reliability is improved (i.e., to minimize the outage due to random blockages that appear during the data transmission phase).
%
As an example, let the set of \acp{RRU} that are used to serve $k$th \ac{UE} with \ac{RRU} indices $\mathcal{B}_k = \{1,2,3\}$. Then, with the assumption of at least $L_k(t) = 2$ available links, the serving set of unblocked \acp{RRU} available to $k$th \ac{UE} can be any one of following combinations:
\begin{equation}
\label{eq:RRU-indices}
    \widehat{\mathcal{B}}_k(L_k(t)) = \bigl\{ \{1,2\}, \{1,3\}, \{2,3\}, \{1,2,3\} \bigr\}.
\end{equation}

Let $C(L_k(t))$ denote the cardinality of set $\widehat{\mathcal{B}}_k(L_k(t))$, defined as $C(L_k(t))\!=\!\sum_{l=L_k(t)}^{|{\mathcal{B}}_k|} \frac{|{\mathcal{B}}_k|!}{l! \ (|{\mathcal{B}}_k| - l)!}$. We use $\mathcal{B}_k\comb$ to represent the $c$-th subset of $\widehat{\mathcal{B}}_k(L_k(t))$, i.e., subset $\mathcal{B}_k\comb \in \widehat{\mathcal{B}}_k(L_k(t))$ such that cardinality $|\mathcal{B}_k\comb| \geq L_k(t)$ for all $c = 1,\ldots, C(L_k(t))$ during time slot~$t$. Then, the received \ac{SINR} of $k$th \ac{UE} for $\mathcal{B}_k\comb$  (i.e., $c$-th subset) is obtained by excluding the potentially blocked \acp{RRU} in expression~\eqref{eq:SINR}, and it can be expressed as
\begin{equation}
\label{eq:SINR-Comb}
    \Gamma_k(\mathbf{F}(t), \mathcal{B}_k\comb) = \frac{\Bigl| {\textstyle \sum\limits_{b\in \mathcal{B}_k\comb}} \mathbf{h}_{b,k}\herm(t) \mathbf{f}_{b,k}(t) \Bigr|^2  }{\sigma_k^2  +  {\textstyle \sum\limits_{u \in \mathcal{K} \backslash k}} \Bigl| {\textstyle \sum\limits_{g\in \mathcal{B}_{u} \backslash \mathcal{D}_k\comb}} \mathbf{h}_{g,k}\herm(t) \mathbf{f}_{g,u}(t) \Bigr|^2 } ,
\end{equation}
where $\mathcal{D}_k\comb = \mathcal{B}_k \backslash \mathcal{B}_k\comb$ denotes a subset of potentially blocked \acp{RRU} which are excluded from the interfering links to $k$th~\ac{UE}. 
Thus, the pessimistic  achievable rate\footnote{We consider Gaussian signalling and each \ac{UE} decodes its intended signal by treating all other interfering signals as noise~\eqref{eq:SINR-Comb}.} 
for $k$th~\ac{UE} during time slot~$t$ can be expressed as 
\begin{equation}
    \label{eq:rate}
    r_k(t) = \log_2\bigl(1+\gamma_k(t)\bigr),
\end{equation}
where $\gamma_k(t) \! = \! \underset{c}{\mathrm{min}} \bigl( \Gamma_k(\mathbf{F}(t), \mathcal{B}_k\comb) \bigr), \ \forall c\! = \! 1,\ldots, C({L_k}(t)), \ \forall k$. In practice, the adverse channel condition and signaling overhead limits the maximum number of cooperating \acp{RRU} for each user (i.e., $\mathcal{B}_k \ \forall k$)~\cite{CRAN-2015}. Thus, the subset combinations $C({L_k}(t))$ are fairly small for modestly sized systems~\cite{Dileep_TWC2021}.

It should be noted that each \ac{UE}~$k$ still coherently receives the signal from all $\mathcal{B}_k$, unless \acp{RRU} are not available during the downlink data transmission phase due to random blockage. However, the actual \acp{RRU} available to serve $k$th \ac{UE} cannot be known a priori in the dynamic blockage environment. Therefore, to maintain a reliable connectivity at each time slot, BBU preemptively underestimate the achievable \ac{SINR} over all possible subset combinations $\mathcal{B}_k\comb\!\!~\in~\!\!\widehat{\mathcal{B}}_k(L_k(t))$, which is associated with the design parameter $L_k(t) \ \forall k$, as it will become clear in Section~\ref{subsec:DynSubSet_Selection}.

%% ----- %%%% ----- %%%% ----- %%%% ----- %%

\subsection{Network Queueing  Model}
\label{subsec:Traffic_Model}
We assume that the \ac{BBU} maintains a set of internal queues for storing network layer data of all \acp{UE}~\cite[Ch. 5]{neely2010stochastic}. Let $Q_k(t)$ denote the current queue backlog of $k$th~UE during time slot~$t$, and $A_k(t)$ represents the amount of data that exogenously arrive to it, with the mean arrival rate of $\mathbb{E}[A_k(t)] = {\lambda}_k$. Then the dynamics of queue $Q_k(t)$ can be expressed as
\begin{equation}
\label{eq:Queue_Dynamics}
    Q_k(t+1) = \bigl[Q_k(t) - r_k(t) + A_k(t)\bigr]^{+}, \ \  \forall k \in \mathcal{K},
\end{equation}
where $r_k(t)$ is transmission rate defined in expression~\eqref{eq:rate}. Furthermore, let $\overline{Q}_k$ denote the time-averaged queue associated with $k$th \ac{UE}, defined as
\begin{equation}
    \label{eq:Queue-Markov}
     \overline{Q}_k \triangleq  \lim_{T \to \infty} \frac{1}{T} {\textstyle \sum\limits_{t=0}^{T-1}} \mathbb{E}\bigl[Q_k(t)\bigr],
\end{equation}
where the expectation depends on the control policy, and is with respect to the random channel states and arrivals. 
%%%%%Here, we adopt the notion of \emph{strong stability}, and we say that the network is strongly stable if~[Neely]

According to the Little's law, the average delay is directly proportional to the average queue length $\overline{Q}_k$~\cite[Ch. 1.4]{gross2008fundamentals}. Hence, for $k$th \ac{UE}, we can achieve the desired latency requirements by imposing a constraint on its queue length at each time slot. Here, we use a probabilistic constraint on the queue length, which is defined~as 
\begin{equation}
    \label{eq:Prob-Queue-Const}
    \mathrm{Pr}\bigl\{ Q_k(t) \geq Q_{k}^{\mathrm{th}} \bigr\} \leq \epsilon, \ \ \forall t,
\end{equation}
where $Q_k^{\mathrm{th}}$ is the maximum allowable queue length for $k$th \ac{UE} and $\epsilon\! \ll\! 1$ is the tolerable queue length violation probability.

%% ----- %%%% ----- %%%% ----- %%%% ----- %%
%
\subsection{Problem Formulation}
\label{subsec:problem}
Our objective is to develop a power-efficient and reliable downlink transmission strategy for C-RAN based dynamic \ac{mmWave} system, while satisfying the user-specific latency requirements. Specifically, we consider a problem of time-average sum-power minimization for \ac{mmWave} communication with random channel blockages, subject to the maximum allowable queue length constraint for each \ac{UE}, expressed as
\begin{subequations}
\label{eq:P1-prob}
\begin{align}
	\displaystyle 
	 \underset{\mathbf{F}(t), {\gamma}_k(t), \ \forall t}{\mathrm{min}} \quad & \lim_{T \to \infty} \frac{1}{T} {\textstyle \sum\limits_{t=0}^{T-1}} \biggl( {\textstyle \sum\limits_{b\in\mathcal{B}} \sum\limits_{k \in \mathcal{K}}} \mathbb{E}\bigl[\|\mathbf{f}_{b,k}(t)\|^2\bigr] \biggr) \label{eq:P1} \\
	 \mathrm{s. t.} \quad
	&\displaystyle  \mathrm{Pr}\bigl\{ Q_k(t) \geq Q_{k}^{\mathrm{th}} \bigr\} \leq \epsilon, \ \ \forall k \in \mathcal{K}, \ \forall t \label{eq:P1-C2} \\
	\begin{split}
	&\displaystyle \gamma_k(t) = \underset{c}{\mathrm{min}} \bigl( \Gamma_k(\mathbf{F}(t), \mathcal{B}_k\comb) \bigr), \\ 
	& \quad  \forall c = 1,\ldots, C(L_k(t)), \ \forall k \in \mathcal{K}, \ \forall t, \end{split} \label{eq:P1-C1} 
	%
%%	\begin{split}
%%	&\displaystyle \gamma_k(t) = \underset{c = 1,\ldots, C(L_k)}{\text{min}} \Gamma_k(\mathbf{F}(t), \mathcal{B}_k\comb), %%\\ 
	%%& \qquad \quad \quad 
%% \ \	\forall k \in \mathcal{K}, \ \forall t, 	\end{split} \label{eq:P1-C1}
\end{align}
\end{subequations}
where the function $\Gamma_k(\mathbf{F}(t), \mathcal{B}_k\comb)$ is defined in~\eqref{eq:SINR-Comb}. The constraint~\eqref{eq:P1-C2} ensures that the queue backlog of each user is less than $Q_{k}^{\mathrm{th}}$ at each time slot with the probability $1-\epsilon$, and thus ensures the desired probabilistic latency requirements.
Note that for each user $k$, the constraint~\eqref{eq:P1-C1} is a pessimistic estimate of achievable \ac{SINR}. More specifically, for a given parameter $L_k(t)$, \ac{BBU} models the \ac{SINR} of $k$th \ac{UE} over all possible subset combinations of potentially available \acp{RRU}~$\mathcal{B}_k\comb$ from the serving set $\widehat{\mathcal{B}}_k(L_k(t))$ (see Section~\ref{subsec:Blockage}). Then, we allow BBU to proactively use the pessimistic estimate of \ac{SINR} in order to allocate the downlink rate for the users such that transmission reliability is improved (i.e., to minimize the outage due to random blockages that appear during the transmission phase).
%

%% -------------------------------------- Beamformer Design -------------------------------------- %%

\section{Dynamic Algorithm via Lyapunov Optimization}
\label{sec:DynamicAlgo_Lyapunov}

The problem~\eqref{eq:P1-prob} is intractable as it consists of a long-term time-average sum-power objective function~\eqref{eq:P1}, non-linear probabilistic queue length constraint~\eqref{eq:P1-C2}, and a large number of coupled non-convex \ac{SINR} expressions~\eqref{eq:P1-C1}. %Hence, problem~\eqref{eq:P1-prob} is intractable. 
In this section, we handle the first two sources of intractability, and derive a dynamic control algorithm for~\eqref{eq:P1-prob} by using the Lyapunov optimization technique~\cite{neely2010stochastic}. The proposed convex relaxation via \ac{FP} techniques and the closed-form iterative algorithms are then provided in Section~\ref{subsec:SINR_Approximation}.

We start by transforming the probabilistic queue length constraint~\eqref{eq:P1-C2} into a time-average constraint using the Markov's inequality, i.e., $ \mathrm{Pr}\{ Q_k \geq Q_{k}^{\mathrm{th}} \} \le \mathbb{E}[Q_k]/Q_{k}^{\mathrm{th}}, \ \forall k$ {\cite{gross2008fundamentals}}. Thereby, problem~\eqref{eq:P1-prob} can be rewritten~as
\begin{subequations}
\label{eq:mP1-prob}
\begin{align}
	\displaystyle 
	 \hspace{-1mm} \underset{\mathbf{F}(t), {\gamma}_k(t), \ \forall t}{\mathrm{min}} \  & \lim_{T \to \infty} \frac{1}{T} {\textstyle \sum\limits_{t=0}^{T-1}} \biggl( {\textstyle \sum\limits_{b\in\mathcal{B}} \sum\limits_{k \in \mathcal{K}}} \mathbb{E}\bigl[\|\mathbf{f}_{b,k}(t)\|^2\bigr] \biggr) \label{eq:mP1} \\
	 \mathrm{s. t.} \ 
    \begin{split}
	&\displaystyle  \lim_{T \to \infty} \! \frac{1}{T} \! {\textstyle \sum\limits_{t=0}^{T-1}} \! \mathbb{E} \bigl[Q_k(t)\bigr] \! \leq \! \epsilon Q_k^{\mathrm{th}}, \ \forall k \in \mathcal{K}, \ \forall t
	\end{split}  \label{eq:mP1-C2} \\
	\begin{split}
	&\displaystyle \gamma_k(t) \le \Gamma_k(\mathbf{F}(t), \mathcal{B}_k\comb) , \\ 
	& \qquad \  \forall c = 1,\ldots, C(L_k(t)), \ \forall k \in \mathcal{K}, \ \forall t, \end{split} \label{eq:mP1-C1}
\end{align}
\end{subequations}
Note that we have relaxed~\eqref{eq:P1-C1} while writing constraint~\eqref{eq:mP1-C1}, and both these constraints are equivalent at the optimality.
%%%% We replace~\eqref{eq:P1-C2} with its linear equivalence~\eqref{eq:Queue-Markov} to render a tractable optimization problem.

Now, we use the Lyapunov optimization framework, specifically, a \emph{drift-plus-penalty} method~\cite{neely2010stochastic} to find a solution of problem~\eqref{eq:mP1-prob}. We enforce the long-term time-average constraint~\eqref{eq:mP1-C2} by transforming it into a queue stability problem~\cite[Ch. 5]{neely2010stochastic}. Specifically, a virtual queue associated with~\eqref{eq:mP1-C2} for each user~$k$ is introduced, and the stability of these virtual queue implies that the constraint~\eqref{eq:mP1-C2} is met.

Let $Z_k(t)$ be the virtual queue associated with~\eqref{eq:mP1-C2} for $k$th~\ac{UE}, and we update $Z_k(t)$ as
\begin{equation}
    \label{eq:Virtual-Queue}
    Z_k(t+1) = \bigl[Z_k(t) + Q_k(t+1) - \epsilon Q_k^{\mathrm{th}} \bigr]^{+},  \ \ \forall k\in\mathcal{K}.
\end{equation}
The expression~\eqref{eq:Virtual-Queue} can be interpreted as a queue dynamics for $k$th UE with arrival rate $Q_k(t+1)$ and service rate $\epsilon Q_k^{\mathrm{th}}$. It can be observed from~\eqref{eq:Virtual-Queue} that if the queue length of a user is larger than the delay tolerance, the virtual queue will increase. Therefore, if the virtual queues $\{Z_k(t)\}_{k\in\mathcal{K}}$ are stable, then by using~\cite[Theorem 2.5]{neely2010stochastic} we can show that~\eqref{eq:mP1-C2} is satisfied.

We now define Lyapunov function and its drift, which is used to stabilize queues $\{Z_k(t)\}_{k\in\mathcal{K}}$. For a compact representation, let ${\mathbf \Psi}(t)=[Z_1(t),\ldots, Z_K(t), Q_1(t), \ldots, Q_K(t)]\tran$ denote a collection of virtual and actual queues. Then we define following quadratic Lyapunov function:
\begin{equation}
\label{eq:Lyp-fn}
 \mathcal{L}( {\mathbf \Psi}(t) ) \triangleq \frac{1}{2}  {\textstyle \sum\limits_{k\in\mathcal{K}}} Z_k(t)^2.
\end{equation} 
Intuitively, we can observe that if $\mathcal{L}({\mathbf \Psi}(t))$ is small, then all queues $\{Z_k(t)\}_{k\in\mathcal{K}}$ are small. Contrarily, if $\mathcal{L}( {\mathbf \Psi}(t) )$ is large then at least one of the queues is large. 
Thus, by minimizing a drift of $\mathcal{L}({\mathbf \Psi}(t))$ from one time slot to another, queues $\{Z_k(t)\}_{k\in\mathcal{K}}$ can be stabilized. 

%%The expected drift of Laypunov function $\mathcal{L}({\mathbf \Psi}(t))$ can be expressed as~\cite{neely2010stochastic} 
Then the Lyapunov drift~\cite[Ch. 5]{neely2010stochastic}, which describes the change in network congestion between consecutive time slots, can be expressed~as
\begin{align}
    \label{eq:Lyapunov-drif-I}
    \triangle ({\mathbf \Psi}(t))  & =  \mathbb{E}\bigl[\mathcal{L}({\mathbf \Psi}(t+1)) - \mathcal{L}({\mathbf \Psi}(t)) | {\mathbf \Psi}(t) \bigr] \nonumber \\
                                   & =  \frac{1}{2}  \mathbb{E} \Bigl[ {\textstyle \sum\limits_{k\in\mathcal{K}}} \bigl( Z_k(t+1)^2 - Z_k(t)^2 \bigr) \big|  {\mathbf \Psi}(t) \Bigr].
\end{align}    
Next, by using expressions~\eqref{eq:Queue_Dynamics} and \eqref{eq:Virtual-Queue} in~\eqref{eq:Lyapunov-drif-I}, an upper bound of drift $\triangle ({\mathbf \Psi}(t))$ can be expressed~as\footnote{To obtain~\eqref{eq:Lyapunov-drif}, we have used the fact that $([a + b - c]^{+})^2 \leq (a + b - c)^2$ for any $a\geq 0$, $b\geq0$, and $c\ge 0$.} %%%%(detailed derivation is omitted due to space limitation)
\begin{align}
    \label{eq:Lyapunov-drif}
    \triangle ({\mathbf \Psi}(t)) & \leq  \zeta +\Phi (t)  -  \\ & \quad \mathbb{E}\Bigl[ {\textstyle \sum\limits_{k\in\mathcal{K}}} (Q_k(t) + A_k(t) + Z_k(t))r_k(t) \big| {\mathbf \Psi}(t) \Bigr]\nonumber, %%%%\nonumber 
\end{align}
where $\zeta$ and $\Phi (t)$  are positive constants, and satisfy the following condition\footnote{We have assumed that second moments of arrival and transmission processes are bounded~\cite{neely2010stochastic}.
The derivation is omitted due to lack of space, and we refer the reader to~\cite{neely2010stochastic} for the details.
} for all time~slots:
\begin{align*}
    \zeta  \ge  & \frac{1}{2} \mathbb{E} \Bigl[ {\textstyle \sum\limits_{k\in\mathcal{K}}} A_k(t)^2 +  {r}_k(t)^2  \big| {\mathbf \Psi}(t) \Bigr], \\
    \Phi (t) = &  {\textstyle \sum\limits_{k\in\mathcal{K}}} \Bigl[  \frac{1}{2} (\epsilon Q_k^{\mathrm{th}})^2 + \frac{1}{2} Q_k(t)^2 \\ & \qquad + Z_k(t)Q_k(t) + \bigl(Q_k(t) + Z_k(t)\bigr){A}_k(t) \Bigr].
\end{align*}

Now we define following \emph{drift-plus-penalty} function~\cite{neely2010stochastic} for problem~\eqref{eq:mP1-prob}:
\begin{align}
    \label{eq:drift-penality}
     &\triangle ({\mathbf \Psi}(t)) + V \mathbb{E} \Bigl[ {\textstyle \sum\limits_{b\in\mathcal{B}} \sum\limits_{k \in \mathcal{K}}}    \|\mathbf{f}_{b,k}(t)\|^2  \big| {\mathbf \Psi}(t) \Bigr], %%\nonumber
     %%\\
     %%&  \qquad \quad \leq \mathbb{E}\Bigl[ V {\textstyle \sum\limits_{b\in\mathcal{B}} \sum\limits_{k \in \mathcal{K}}}    \|\mathbf{f}_{b,k}(t)\|^2 + {\textstyle \sum\limits_{k\in\mathcal{K}}} \bigl(Q_k(t) + A_k(t) + Z_k(t)\bigr)r_k(t) \big| \Psi(t) \Bigr],
\end{align}
where $V \geq 0$ is a trade-off parameter. By using expression~\eqref{eq:Lyapunov-drif} in~\eqref{eq:drift-penality}, and 
%%%%%Therefore, by 
minimizing the upper bound of~\eqref{eq:drift-penality} subject to constraint~\eqref{eq:mP1-C1} at each time slot, we can stabilize queues $\{Z_k(t)\}_{k\in\mathcal{K}}$ and minimize the sum power objective function of problem~\eqref{eq:mP1-prob}.   
Thus, we utilize the concept of \emph{opportunistic minimization of an expectation}~\cite[Ch. 1.8]{neely2010stochastic} to minimize the drift-plus-penalty function~\eqref{eq:drift-penality}, and obtain a dynamic control algorithm as detailed in Algorithm~\ref{algLyapunov}. 
%
%
%% ----- %%%% ----- %%
\vspace*{-5pt}
\SetArgSty{textnormal}
\begin{algorithm}[]
	\caption{Dynamic control algorithm for~\eqref{eq:P1-prob}} %%%
	\label{algLyapunov}
	\SetAlgoLined
	For a given time slot~$t$, observe  current queue backlogs $\big\{{Q}_k(t), \ {Z}_k(t)\big\}$ and solve following problem:
\begin{subequations}
\label{eq:P1-prob-2}
\begin{align}
	\displaystyle
	\begin{split}
	 \hspace{-4mm} \underset{\mathbf{F}(t), {\gamma}_k(t)}{\mathrm{min}} \quad & V {\textstyle \sum\limits_{b\in\mathcal{B}} \sum\limits_{k \in \mathcal{K}}} \|\mathbf{f}_{b,k}(t)\|^2    
	 -{\textstyle \sum\limits_{k\in\mathcal{K}}}  \bigl( Q_k(t)  \\ & \qquad +  A_k(t) + Z_k(t) \bigr)\log_{2}\bigl(1+\gamma_k(t)\bigr)
	\end{split}
	 \label{eq:P1-2} \\
	 \mathrm{s. t.} \quad
	\begin{split}
	&\displaystyle \gamma_k(t) \le \Gamma_k(\mathbf{F}(t), \mathcal{B}_k\comb),  \ \  \forall c , \ \forall k \in \mathcal{K}. 	
	\end{split} \label{eq:P1-C1-2}
\end{align}
\end{subequations}
    \\
    Update queues ${Q}_k(t+1)$ and ${Z}_k(t+1)$ by using \eqref{eq:Queue_Dynamics} and \eqref{eq:Virtual-Queue}, respectively, for all $k\in\mathcal{K}$ \\
    Set $t = t +1$, and go to step~$1$
\end{algorithm}
\vspace*{-8pt}
%% ----- %%%% ----- %%
%

At each time slot of Algorithm~\ref{algLyapunov}, we need to solve problem~\eqref{eq:P1-prob-2} to find optimal beamforming vectors $\mathbf{F}(t)$. Therefore, we derive iterative algorithms for this in the next section.

%%%%%The coupled and non-convex constraints are approximated with the sequence of convex subset by using two alternative approaches, namely the \ac{SCA} framework~\cite{beck2010sequential} and the \ac{FP} techniques~\cite{FP_Shen_2018}. Both approaches allow an implementation of iterative low-complexity beamformer design with distinct convergence characteristics that enable tailored complexity and convergence performance depending on the specific implementations.

%% -------------------------------------- Iterative Algorithm -------------------------------------- %%
%
%
\section{Iterative Algorithms for Problem~\eqref{eq:P1-prob-2}}
\label{subsec:SINR_Approximation}

The problem~\eqref{eq:P1-prob-2} is intractable as-is, mainly due to coupled and non-convex \ac{SINR} expressions~\eqref{eq:P1-C1-2}. %%%%Several approaches have been outlined in the existing literature to handle the \ac{SINR} non-convexity. 
In this section, we elaborate on finding the solution of problem~\eqref{eq:P1-prob-2} by using the  \ac{FP} {quadratic transform} techniques~\cite{FP_Shen_2018}. 
%%%%%%%%However, the update rules in \ac{FP} techniques are given in a closed-form expressions, and it is proven to attain the solution at each iteration~\cite{FP_Shen_2018}. %%%% Moreover, the \ac{FP} method provides comparatively faster convergence.  
%%%%On the contrary, the \ac{SCA} provides comparatively faster and monotonic convergence for  problem~\eqref{eq:P1-prob-2}. 
%
%%%%%%%%%%Both approaches allow an implementation of beamformer design with distinct convergence characteristics, which enable tailored complexity and processing performance (depending on the application specific requirements).
%
In addition, we also provide a low-complexity  beamformer design via iterative evaluation of the closed-form \ac{KKT} optimality conditions and by the parallel update of beamformers corresponding to the spatially distributed~\acp{RRU}.

It is worth highlighting, in the \ac{FP} techniques~\cite{FP_Shen_2018, FP_Shen_2018_2, CoMP-IoTJ_2021}, all non-convex constraints are approximated with the sequence of convex subset, and the underlying convex subproblem is then iteratively solved until the convergence of objective function to a local solution. 
The \ac{FP} {quadratic transform} based solutions have been widely used in many applications, e.g.,  power control, energy efficiency, and multi-antenna interference coordination~\cite{FP_Shen_2018, FP_Shen_2018_2, CoMP-IoTJ_2021}. For example, the  SINR relaxation via \ac{FP} techniques is provided for downlink in~\cite[Section IV]{FP_Shen_2018}, \cite[Section III]{CoMP-IoTJ_2021} and for uplink in~\cite[Section V]{FP_Shen_2018_2}, assuming perfect CSI and no blockages. %%%%%In~\cite{Dileep_TWC2021}, we provide reliable transmission schemes in the presence of blockages. 
Further, all these algorithms are studied for the static case (i.e., the resource allocation problem for a given instance is studied). %%% without considering the dynamics of the network.
In view of the prior works, there lacks a systemic approach for the design of  beamforming vectors in the {JT-CoMP} scenarios, while accounting for the uncertainties of mmWave radio channel in a time-average dynamic network with the stringent user-specific latency and reliability requirements, thus motivating the current work. 
%
%%%%%%%%We extend these approaches to take into consideration coherent multi-point transmission and provide a novel grouping of a multitude of potentially coupled and non-convex SINR conditions that raise from the link blockage subsets.

%% ----- %%%% ----- %%%% ----- %%%% ----- %%

\subsection{Solution via Fractional Program (FP) techniques}
\label{subsec:FracProg_Algorithm}

We start by using the expression of $\Gamma_k(\mathbf{F}, \mathcal{B}_k\comb)$ (see~\eqref{eq:SINR-Comb}), and compactly rewrite~\eqref{eq:P1-C1-2}~as
\begin{equation}
\label{eq:SINR-Den-Num}
    {\gamma}_k(t) \leq \frac{ | \mathbf{h}_{k}\cherm(t) \mathbf{f}_{k}(t) |^2}{\sigma_k^2  + {\textstyle \sum\limits_{u \in \mathcal{K} \backslash k}} \bigl| \mathbf{h}_k\cherm(t) \mathbf{f}_{u}(t)  \bigr|^2},
\end{equation}
%
%
%%%%where ${\mathbf{f}}_{k}(t)\!\! \triangleq \!\! [\mathbbm{1}_{\mathcal{B}_k }(1)\mathbf{f}_{1,k}\tran(t), \ldots, \mathbbm{1}_{ \mathcal{B}_k}(B)\mathbf{f}_{B,k}\tran(t)]\tran \! \in \! \mathbb{C}^{ | {\mathcal{B}} | N \times 1}$ denotes the stacked transmit beamformer and ${\mathbf{h}}_{k}\comb(t) \triangleq [\mathbbm{1}_{\mathcal{G}_k\comb}(1)\mathbf{h}_{1,k}\tran(t), \ldots,  \mathbbm{1}_{\mathcal{G}_k\comb}(B)\mathbf{h}_{B,k}\tran(t)]\tran \in \mathbb{C}^{ | {\mathcal{B}} | N \times 1}$ represents the stacked channel vector. 
%
%
where ${\mathbf{f}}_{k}(t)\!\! \triangleq \!\! [\mathbbm{1}_{\mathcal{B}_k }(1)\mathbf{f}_{1,k}\tran(t), \ldots, \mathbbm{1}_{ \mathcal{B}_k}(B)\mathbf{f}_{B,k}\tran(t)]\tran \! \in \! \mathbb{C}^{ | {\mathcal{B}} | N \times 1}$ and 
${\mathbf{h}}_{k}\comb(t) \!\! \triangleq \!\! [\mathbbm{1}_{\mathcal{G}_k\comb}(1)\mathbf{h}_{1,k}\tran(t), \ldots,  \mathbbm{1}_{\mathcal{G}_k\comb}(B)\mathbf{h}_{B,k}\tran(t)]\tran\!  \in \! \mathbb{C}^{ | {\mathcal{B}} | N \times 1}$  
denotes the stacked beamformer and channel, respectively.
%
%
%%%%%%%%%%%%%%%%%%%%%%%%%%%%%%%%%%%%%%%%%%%%%%%%%%%%%%%%%%%%%%%%%%%%%%%%%%
\begin{comment}
%
where $\mathbf{f}_{k}(t) \in \mathbb{C}^{ | {\mathcal{B}} | N \times 1}$ is the stacked transmit beamformer and 
$\mathbf{h}_{k}\comb(t) \in \mathbb{C}^{ | {\mathcal{B}} | N \times 1}$ is the stacked channel vector, defined as
%
\begin{align*}
    {\mathbf{f}}_{k}(t)  & \triangleq  \big[\mathbbm{1}_{\mathcal{B}_k }(1)\mathbf{f}_{1,k}\tran(t), \ldots, \mathbbm{1}_{\mathcal{B}_k }(b)\mathbf{f}_{b,k}\tran(t), \ldots, \mathbbm{1}_{ \mathcal{B}_k}(B)\mathbf{f}_{B,k}\tran(t) \big]\tran, \\
    %
    {\mathbf{h}}_{k}\comb(t) & \triangleq \big[\mathbbm{1}_{\mathcal{G}_k\comb}(1)\mathbf{h}_{1,k}\tran(t), \ldots, \mathbbm{1}_{\mathcal{G}_k\comb}(b)\mathbf{h}_{b,k}\tran(t), \ldots, \mathbbm{1}_{\mathcal{G}_k\comb}(B)\mathbf{h}_{B,k}\tran(t) \big]\tran.
\end{align*}
%
\end{comment}
%%%%%%%%%%%%%%%%%%%%%%%%%%%%%%%%%%%%%%%%%%%%%%%%%%%%%%%%%%%%%%%%%%%%%%%%%%
%
The indicator function $\mathbbm{1}_{\mathcal{G}_k\comb}(b)$ and $\mathbbm{1}_{\mathcal{B}_j}(b)$ are defined~as
\begin{eqnarray*}
%
%%%\begin{equation*}
\mathbbm{1}_{\mathcal{B}_k }(b) & = & \left\{
    \begin{array}{ll}
        {1} & \text{ if and only if } \ \   b \in \mathcal{B}_k \\
        {0} & \text{ otherwise. }
    \end{array}
\right.
%%%\end{equation*}
\\
%%%\begin{equation*}
\mathbbm{1}_{\mathcal{G}_k\comb}(b) & = & \left\{
    \begin{array}{ll}
        {1} & \text{ if and only if } \ \  b \in \mathcal{B}\setminus\mathcal{D}_k\comb \\
        {0} & \text{ otherwise, }
    \end{array}
\right.
%%%\end{equation*}
%
\end{eqnarray*}
where ${\mathcal{G}_k\comb} = {\mathcal{B} \backslash \mathcal{D}_k\comb}$ for all $c=1,\ldots, C(L_k)$ and $k\in\mathcal{K}$. In this section, we omit time index~$t$ to simplify the notation.

%%A step towards doing so is presented in [6], where we provide reliable  transmission  schemes,  by  preemptively  underestimating  the  achievable  rates  over  the potential link  blockage combinations, in  the presence  of blockages.

%%%%To begin with, l
Let us now examine the characteristic of the objective function in problem~\eqref{eq:P1-prob-2}. We can observe that the right-hand-side~(RHS) of~\eqref{eq:P1-C1-2} is a typical function of multiple fractional parameters (i.e., \ac{SINR} expressions~\eqref{eq:SINR-Den-Num}). Thus, problem~\eqref{eq:P1-prob-2} can be recast as a multi-ratio fractional program problem. Motivated by the findings in~\cite{FP_Shen_2018}, here we adopt the recently proposed   \ac{FP} {quadratic transform} techniques, wherein the non-convex problem is recast as a sequence of convex subproblem, and then iteratively solved until the convergence of objective function. We extend the approaches~\cite{FP_Shen_2018} to take into consideration coherent multi-point transmission and provide a novel grouping of a multitude of potentially coupled and non-convex SINR conditions, that raise from the link blockage subset combinations. We develop the following proposition based on the  \ac{FP} techniques~\cite[Theorem 1]{FP_Shen_2018}.

\noindent \textbf{Proposition 1.} The fractional terms in RHS of~\eqref{eq:SINR-Den-Num}
\begin{equation}
\label{eq:FP-eq1}
    \frac{ | \mathbf{h}_{k}\cherm \mathbf{f}_{k}|^2}{\sigma_k^2  +  {\textstyle \sum\limits_{u \in \mathcal{K} \backslash k}} \bigl| \mathbf{h}_k\cherm \mathbf{f}_{u}  \bigr|^2},
\end{equation}
is equivalent to
\begin{equation}
\label{eq:SINR-FP}
2\Re\bigl\{ \nu_{k,c}^{*} \mathbf{h}_{k}\cherm \mathbf{f}_{k}  \bigr\} - \nu_{k,c}^{*} \Bigl[ \sigma_k^2  + {\textstyle \sum\limits_{u \in \mathcal{K} \backslash k}} \bigl| \mathbf{h}_k\cherm \mathbf{f}_{u} \bigr|^2 \Bigr] \nu_{k,c},
\end{equation}
when the auxiliary variable $\{\nu_{k,c}\}$ has the optimal value as
\begin{equation}
\label{eq:FP-nu}
\nu_{k,c}^{\scriptsize (\star) } = \frac{ \mathbf{h}_{k}\cherm \mathbf{f}_{k}}{\sigma_k^2  + \sum\limits_{u \in \mathcal{K} \backslash k} \bigl| \mathbf{h}_k\cherm \mathbf{f}_{u}  \bigr|^2},
\end{equation}
for all $c\!=\!1,\ldots,C(L_k)$ and $k\in\mathcal{K}$. %%%We refer the reader to \cite[Section IV]{FP_Shen_2018} for the proof and details. 
  We refere the reader to~\cite{FP_Shen_2018} for details. Specifically, the Proposition 1 can be proved by following the steps in \cite[Section IV]{FP_Shen_2018}. 

Thereby using Proposition~1, we can obtain a solution for problem~\eqref{eq:P1-prob-2} 
%%%% by iteratively optimizing the primal variables~$\{\mathbf{f}_k, \gamma_k\}$ and the auxiliary variables $\{\nu_{k,c}\}$, e.g., 
by iteratively solving a sequence of convex subproblems~\cite{FP_Shen_2018}. For example, the convex subproblem for $i$-th iteration along with corresponding dual variables can be expressed~as
\begin{subequations}
\label{eq:P1-FP}
\begin{align}
	\displaystyle 
	\begin{split}
	 \underset{\mathbf{F}, \gamma_k}{\mathrm{min}} \ \ & V {\textstyle\sum\limits_{k \in \mathcal{K}}}   \|\mathbf{{f}}_{k}\|^2 \! \! - \! \!  {\textstyle\sum\limits_{k\in\mathcal{K}}}  (Q_k + A_k + Z_k)\log_{2}(1+\gamma_k)
	 \end{split} \\
	 \mathrm{s. t.} \ \
	\begin{split}
    & \displaystyle 
    e_{k,c} : {\gamma}_k \leq 2\Re\bigl\{ \nu_{k,c}\xiLoopComp \mathbf{h}_{k}\cherm \mathbf{f}_{k}  \bigr\}   
    - \nu_{k,c}\xiLoopComp \Bigl[ \sigma_k^2   + \\ & \qquad \qquad {\textstyle\sum\limits_{u \in \mathcal{K} \backslash k}} \bigl| \mathbf{h}_k\cherm \mathbf{f}_{u} \bigr|^2 \Bigr] \nu_{k,c}\xiLoop,  \ \ \forall c, \ \forall k \in \mathcal{K},
    \end{split}
	\label{eq:P1-C1-FP}
\end{align}
\end{subequations}
where $\{e_{k,c}\}$ are non-negative Lagrangian multipliers associated with constraints~\eqref{eq:P1-C1-FP}. Note that~\eqref{eq:P1-FP} provides an approximate solution for~\eqref{eq:P1-prob-2} in the vicinity of a fixed operating point $\{\nu_{k,c}\xiLoop\}$. Specifically, for fixed auxiliary variables, we first optimize the primal optimization variables, and then update auxiliary variables with the current solution using expression~\eqref{eq:FP-nu}. 
Thus, %%%similar to \ac{SCA} framework~\cite{beck2010sequential}, 
the proposed \ac{FP} quadratic transform technique is  based on an iterative optimization framework. However, the update rules of the auxiliary variables are given in a closed-form expression~\eqref{eq:FP-nu}, and it is proven to attain the stationary point solution at each iteration~\cite{FP_Shen_2018, FP_Shen_2018_2}. %%%% the  optimum at each iteration~\cite{FP_Shen_2018}.
Hence, by iteratively solving~\eqref{eq:P1-FP} while updating the auxiliary variables $\{\nu_{k,c}\iLoop\}$ with current solution, we can find a best solution for~\eqref{eq:P1-prob-2}, using existing convex optimization toolboxes, such as $\mathrm{CVX}$~\cite{cvx}. The beamformer design with the proposed \ac{FP}  relaxations has been summarized in Algorithm~\ref{algFP}.
%
%
%%%%\vspace{-5pt}
\SetArgSty{textnormal}
\begin{algorithm}[]
	\caption{FP based algorithm for~\eqref{eq:P1-prob-2}}
	\label{algFP}
	\SetAlgoLined
	Set $i = 1$ and initialize with a feasible %%%starting point % choose any feasible initial points 
	$\bigl\{\nu_{k,c}\oLoop\bigr\}$, \ $\forall c, \ \forall k$ \\
	\Repeat{convergence or for fixed number of iterations}{
	{Solve \eqref{eq:P1-FP} with $\bigl\{\mathbf{v}_{k,c}\xiLoop\bigr\}$ and denote the local optimal values as $\bigl\{\mathbf{f}_{k}\iLoop, {\gamma}_k\iLoop\bigr\}$ \\
	Obtain $\nu_{k,c}\iLoop$ using \eqref{eq:FP-nu} with updated $\bigl\{\mathbf{f}_{k}\iLoop \bigr\}$  \\
	Set $i = i+1$
	}
}
\end{algorithm}
%%%%\vspace{-5pt}
%
%

\subsubsection{Solution via KKT Conditions for Problem~\eqref{eq:P1-FP}}
\label{subsec:Approx-subProblem_FP}

We can observe that for fixed auxiliary variables $\{\nu_{k,c}\}$, each subproblem~\eqref{eq:P1-FP} is a convex problem with respect to variables $\{\mathbf{f}_k, \gamma_k\}$. Thus, we can efficiently obtain the solution by the Lagrangian multiplier method. Here, we tackle~\eqref{eq:P1-FP} by iteratively solving a system of \ac{KKT} optimality conditions~\cite[Ch. 5.5]{boyd2004convex}, and, in general, it admits a closed-form solution that does not rely on generic convex solvers.
The Lagrangian $\mathfrak{L}_{\mathrm{FP}}(\mathbf{F}, {\gamma}_k, \nu_{k,c}, e_{k,c})$ of problem~\eqref{eq:P1-FP}  is given in~\eqref{eq:Lagrangian-FP}. 
%
%%%%%%%%%%%%%%%%%%%%%%%%%%%%%%%%%%%
%
%
\begin{figure*}[t]
%%%%\hrulefill
%%%%%%\vspace*{-1pt}
% ensure that we have normalsize text
\normalsize
% Store the current equation number.
\setcounter{mytempeqncnt}{\value{equation}}
% Set the equation number to one less than the one
% desired for the first equation here.
% The value here will have to changed if equations
% are added or removed prior to the place these
% equations are referenced in the main text.
\setcounter{equation}{29}
\begin{align}
\label{eq:Lagrangian-FP}
& \mathfrak{L}_{\mathrm{FP}}(\mathbf{F}, {\gamma}_k, \nu_{k,c} e_{k,c})  = 
{\textstyle \sum\limits_{k=1}^K} \biggl[ V  \|{\mathbf{f}}_{k}\|^2  
    - (Q_k + A_k + Z_k)\log_{2}(1+\gamma_k)
    + {\textstyle \sum\limits_{c=1}^{C({L_k})}} e_{k,c}\gamma_k   \nonumber \\ & \quad \qquad \qquad \qquad \qquad \quad
    - 2 \! {\textstyle \sum\limits_{c=1}^{C({L_k})}} \! e_{k,c} \Re\bigl\{ \nu_{k,c}\xiLoopComp \mathbf{h}_{k}\cherm \mathbf{f}_{k}  \bigr\} \! + \! {\textstyle \sum\limits_{c=1}^{C({L_k})}} \! e_{k,c} |\nu_{k,c}\xiLoop|^2 \sigma_k^2 
    \! + \! {\textstyle \sum\limits_{u \in \mathcal{K} \backslash k}} \! {\textstyle \sum\limits_{c=1}^{C({L_u})}} \! e_{u,c} \nu_{u,c}\xiLoopComp | \mathbf{{h}}_{u}\cherm \mathbf{{f}}_{k} |^2 \nu_{u,c}\xiLoop \biggr].
\end{align}
% Restore the current equation number.
\setcounter{equation}{\value{mytempeqncnt}}
\setcounter{equation}{30}
% IEEE uses as a separator
\hrulefill
% The spacer can be tweaked to stop underfull vboxes.
\vspace*{-12pt}
\end{figure*}
%
%
%%%%%%%%%%%%%%%%%%%%%%%%%%%%%%%%%%%
%
%
The stationary conditions for $k$th \ac{UE} is obtained by differentiating~\eqref{eq:Lagrangian-FP} with respect to associated primal optimization variables $\{\mathbf{{f}}_{k}, {\gamma}_k\}$ for all $k\in\mathcal{K}$~(refer to \cite[Ch. 5.5.3]{boyd2004convex} for details). 
Thus, the stationary conditions of each user~$k$ for problem~\eqref{eq:P1-FP} can be expressed~as
\begin{subequations}
\label{eq:KKT_Derivate_FP}
\begin{align}
\label{eq:Derivate_gamma_FP}
& \nabla_{{\gamma}_k}: {\textstyle \sum\limits_{c=1}^{C({L_k})}} e_{k,c} = \frac{Q_k + A_k + Z_k}{1+{\gamma}_k}, \\
\label{eq:Derivate_Precoder_FP}
& \nabla_{{\mathbf{f}}_k}: {\textstyle \sum\limits_{c=1}^{C(L_k)}} e_{k,c} \Bigl( \nu_{k,c}\xiLoopComp \mathbf{h}_{k}\cherm \Bigr)  
= \mathbf{{f}}_{k}\herm \Bigl( V\mathbf{1} \ + \nonumber \\ & \qquad \qquad \quad {\textstyle \sum\limits_{u \in \mathcal{K} \backslash k} \! \! \sum\limits_{c=1}^{C(L_u)}} e_{u,c} \nu_{u,c}\xiLoopComp \bigl( \mathbf{{h}}_{u}\comb \mathbf{{h}}_{u}\cherm \bigr) \nu_{u,c}\xiLoop \Bigr).   
\end{align}
\end{subequations}
In addition to~\eqref{eq:KKT_Derivate_FP} and primal-dual feasibility constraints, the \ac{KKT} conditions also include the complementary slackness as
\begin{align}
& {e}_{k,c} \biggl\{ {\gamma}_k - 2\Re\bigl\{ \nu_{k,c}\xiLoopComp \mathbf{h}_{k}\cherm \mathbf{f}_{k}  \bigr\} 
+ \nu_{k,c}\xiLoopComp \Bigl[ \sigma_k^2  \ +   \nonumber \\ & \qquad \qquad {\textstyle \sum\limits_{u \in \mathcal{K} \backslash k}} \bigl| \mathbf{h}_k\cherm \mathbf{f}_{u} \bigr|^2 \Bigr] \nu_{k,c}\xiLoop \biggr\} = 0, %%%\nonumber \\ & \qquad \qquad \qquad \qquad \qquad
\ \ \forall c, \ \forall k \in \mathcal{K}. %%=1,\ldots C({L_k})
\label{eq:slackness_ek}
\end{align}

Note that, the user-specific \ac{SINR} constraint~\eqref{eq:P1-C1-FP} is mutually coupled and interdependent over the serving set combinations (see Section~\ref{subsec:Blockage}). Hence, obtaining a closed-form solution for the associated Lagrangian multipliers $\{e_{k,c}\}$ in expression~\eqref{eq:KKT_Derivate_FP} is considerably more difficult than the case with single and non-coupled {QoS} constraint for each user~\cite{Kaleva-DecentralizeJP-2018, CoMP-IoTJ_2021}. Thus, to overcome this challenge, we resort to the subgradient approach, where all non-negative Lagrangian multipliers $\{e_{k,c}\}$ are iteratively solved using the subgradient method~\cite{boyd2003subgradient}. 
The  closed-form steps in the iterative algorithm~are:
% 
%
%%%%\begin{figure*}[htb!]
\begin{subequations}
\label{eq:KKT-i_FP iteration}
\begin{align}
\label{eq:KKT-i_FP, f i}
& \mathbf{{f}}_{k}\iherm   = 
{\textstyle \sum\limits_{c=1}^{C(L_k)}}  e_{k,c}\xiLoop \Bigl( \nu_{k,c}\xiLoopComp \mathbf{h}_{k}\cherm \Bigr) \times \\ \nonumber & \qquad \qquad
\Bigl\{  V\mathbf{1} + \!\!\! {\textstyle \sum\limits_{u \in \mathcal{K} \backslash k} \!\! \sum\limits_{c=1}^{C(L_u)}} \! e_{u,c}\xiLoop \nu_{u,c}\xiLoopComp \bigl(\mathbf{{h}}_{u}\comb \mathbf{{h}}_{u}\cherm \bigr)  \nu_{u,c}\xiLoop \Bigr\}^{-1} \!,  \\ \displaybreak[0]
\label{eq:KKT-i_FP, SINR i}
& {\gamma}_k\iLoop = \frac{Q_k + A_k + Z_k}{ \sum\limits_{c=1}^{C(L_k)} e_{k,c}\xiLoop} - 1, \\ \displaybreak[0]
& \Gamma_k\iLoop(\mathbf{F}, \mathcal{B}_k\comb)  =
\frac{ | \mathbf{{h}}_{k}\cherm \mathbf{{f}}_{k}\iLoop |^2}{\sigma_k^2 + \sum\limits_{u \in \mathcal{K} \backslash k} \bigl| \mathbf{{h}}_{k}\cherm \mathbf{{f}}_{u}\iLoop \bigr| ^2},
\label{eq:KKT-i_FP, gamma_i} \\
%
%
%%%& e_{k,c}\iLoop  = e_{k,c}\xiLoop + \beta \big[ {\gamma}_k\iLoop-\Gamma_k\iLoop(\mathbf{F}, \mathcal{B}_k\comb) \big], \label{eq: KKT-i_FP, ek} \\
&{ e_{k,c}\iLoop  = \Bigl(e_{k,c}\xiLoop + \beta_e \bigl[ {\gamma}_k\iLoop - \text{expression} \ \eqref{eq:SINR-FP}\bigr]  \Bigr)^{+},} \label{eq: KKT-i_FP, ek} \\
& \nu_{k,c}\iLoop  = \frac{\mathbf{h}_{k}\cherm \mathbf{f}_{k}\iLoop}{\sigma_k^2  + \sum\limits_{u \in \mathcal{K} \backslash k} \bigl| \mathbf{h}_k\cherm \mathbf{f}_{u}\iLoop \bigr|^2}. %%%%,
\label{eq: KKT-i_FP, nu} %%%% \\
%
%%%%%%%%%%%%%% & \nu_{k,c}\iLoop = \nu_{k,c}^{\scriptsize (\star)} + \tau (\nu_{k,c}^{\scriptsize (\star)} - \nu_{k,c}\xiLoop) \label{eq: KKT-i_FP, nu_i} 
%
%
\end{align}
\end{subequations}
where $\beta_e$ %%%% and $\tau$ are 
is small positive step-size\footnote{The step size  depends on the system model, as it directly affects the convergence rate and controls the oscillation in the objective function~\cite{boyd2003subgradient}.}. In expression~\eqref{eq: KKT-i_FP, ek}, the dual variables $\{e_{k,c}\}$ are iteratively updated based on the violation of SINR constraint, e.g., the complementary slackness conditions~\eqref{eq:slackness_ek}, using the subgradient method~\cite{boyd2003subgradient}.
%
%%%Moreover, the auxiliary variables~$\{\nu_{k,c}\}$ 
%
The proposed beamformer design by solving a system of closed-form \ac{KKT} expressions is summarized in Algorithm~\ref{algKKT_FP}.
%

%% ----- %%%% ----- %%
\vspace*{-5pt}
\SetArgSty{textnormal}
\begin{algorithm}[]
	\caption{KKT based iterative algorithm for~\eqref{eq:P1-FP}} %%%
	\label{algKKT_FP}
	\SetAlgoLined
	Set $i = 1$ and initialize $\big\{{{v}}_{k,c}\oLoop \big\}$, \ \  
	$\forall c, \ \forall k$    \\ 
	\Repeat{convergence or for fixed iterations}{
	{
	Solve $\mathbf{f}_{k}\iLoop$ from~\eqref{eq:KKT-i_FP, f i} with $\bigl\{\nu_{k,c}\xiLoop, e_{k,c}\xiLoop \bigr\}$ \\ %
	Obtain  ${\gamma}_k\iLoop$ from~\eqref{eq:KKT-i_FP, SINR i} %%%with $\big\{\mathbf{{f}}_{b,k}\xiLoop, {\gamma}_k\xiLoop, a_{k,c}\xiLoop\big\}$   
	\\
	Calculate  ${\Gamma}_k\iLoop(\mathbf{F}, \mathcal{B}_k\comb)$ from~\eqref{eq:KKT-i_FP, gamma_i} with updated $\bigl\{\mathbf{f}_{k}\iLoop\bigr\}$  \\
	Obtain $e_{k,c}\iLoop$ using~\eqref{eq: KKT-i_FP, ek} %%%%with~$\bigl\{{\gamma}_k\iLoop, \Gamma_k\iLoop(\mathbf{F},\mathcal{B}_k\comb)\bigr\}$ 
	\\
	Solve $\nu_{k,c}\iLoop$ from~\eqref{eq: KKT-i_FP, nu} with updated $\bigl\{\mathbf{f}_{k}\iLoop\bigr\}$ \\
	% 
	%%%Update $\nu_{k,c}\iLoop$ using~\eqref{eq: KKT-i_FP, nu_i} \\
	%
	Set $i = i+1$
	}
}
\end{algorithm}
\vspace*{-5pt}
%% ----- %%%% ----- %%

%%%iterative optimization framework
%%\vspace*{-5pt}
\subsection{Feasible Initialization and Complexity Analysis}
\label{subsec:feasible_Initial_Complexity}
\subsubsection{Feasible Initialization}
In the  \ac{FP} methods, the non-convex constraints~\eqref{eq:P1-C1-2} are approximated with the sequence of convex subsets, and then iteratively solved until the convergence of objective function to a stationary point solution. Thus, it is important to initialize the proposed iterative algorithms with a feasible starting point,
%%%%~i.e., $\{\mathbf{f}_k\oLoop, {\gamma}_k\oLoop, \nu_{k,c}\oLoop, a_{k,c}\oLoop, e_{k,c}\oLoop\} \ \forall c = 1,2,\ldots,C(L_k), \ k\in\mathcal{K}$, %
as it impacts the problem feasibility and the rate of convergence~\cite{FP_Shen_2018, FP_Shen_2018_2}.
To this end, one possible option for a feasible $\{\mathbf{f}_k\oLoop \}$ is to use any randomly generated beamforming vector. Then, compute the lower bound of achievable \ac{SINR} from~\eqref{eq:SINR-Comb}, i.e., ${\gamma}_k\oLoop \!=\! \underset{c}{\mathrm{min}}\big( \Gamma_k({\mathcal{B}}_{k}\comb) \big), \forall c\!=\!1,2,\dots,C(L_k), \ \forall k\!\in\!\mathcal{K}$. Furthermore, with the feasible $\{\mathbf{f}_k\oLoop \}$, the initial values of the auxiliary variables~$\{\nu_{k,c}\oLoop\}$ can be computed using~\eqref{eq:FP-nu}. The non-negative Lagrangian multiplier $\{e_{k,c}\oLoop\}$ in Algorithm~\ref{algKKT_FP} are initialized  such that $\sum_{c=1}^{C(L_k)} e_{k,c}\oLoop>0$, e.g., at least one of the coupled \ac{SINR} constraint is active for each user (see \eqref{eq:Derivate_gamma_FP} and \eqref{eq:KKT-i_FP, SINR i} for more details). %%%%Similarly, the non-negative  dual variables $\{e_{k,c}\oLoop\}$ in Algorithm~\ref{algKKT_FP} are initialized randomly such that LHS of~\eqref{eq:Derivate_gamma_FP} is strictly positive. 
It is worth noting that the initialization of the iterative Algorithms with different \emph{feasible} initial values~$\big\{\mathbf{f}_k\oLoop, {\gamma}_k\oLoop, \nu_{k,c}\oLoop, e_{k,c}\oLoop\big\}$, in general,  does not impact the local solution of problem~\eqref{eq:P1-prob-2}, provided a sufficient number of iterations~\cite{boyd2004convex}. For detail on the convergence and the stationary point solution, we refer the readers to~\cite{FP_Shen_2018, FP_Shen_2018_2}.  %%quadratic transform 

\subsubsection{Complexity Analysis}
The approximated convex subproblem~\eqref{eq:P1-FP} can be solved in a generic convex optimization solver, i.e., as a sequence of second-order cone programs (SOCP)~\cite{Lobo-Vandenberghe-Boyd-Lebret-98}. The interior points methods are generally adopted to efficiently solve the SOCP formulations, wherein, the computational complexity of each iteration scales with the length of system wide joint beamforming vectors~$(|\mathcal{B}|N)$ and the number of constraints~\cite{Lobo-Vandenberghe-Boyd-Lebret-98}. In this case, it can be shown that solving each  subproblem~\eqref{eq:P1-FP} requires  $\mathcal{O}\bigl((|\mathcal{B}|N)^{3.5}\bigr)$ arithmetic operations. %%Thus, particularly, for dense mmWave deployments with large $N$ and $B$, the complexity quickly becomes intractable in practice. 
The computational complexity of the iterative algorithms, e.g., by solving a system of closed-form \ac{KKT} optimality conditions, is mainly dominated  by expression~\eqref{eq:KKT-i_FP, f i} for each subproblem~\eqref{eq:P1-FP}. We can observe that expression~\eqref{eq:KKT-i_FP, f i} consists of matrix multiplications and inverse operations, and the computational complexity of each iteration scales with the length of \ac{RRU} specific beamforming vector~$(N)$. Thus, it can be shown that solving each subproblem via iterative evaluation of \ac{KKT} optimality conditions require  $\mathcal{O}\bigl(|\mathcal{B}_k|N^{3}\bigr)$ arithmetic operations.  
Thus, algorithms based on iterative evaluation of closed-form \ac{KKT} optimality conditions provide a conveniently parallel structure for the beamformer design with relatively lower complexity compared to the joint beamformer optimization across all~\acp{RRU}.
%
%%%%%%%% Furthermore, the  complexity of matrix inversion in~\eqref{eq:KKT-i_FP, f i} can be improved by solving $\{\mathbf{f}_k\}$ from linear equations~\cite{boyd2004convex}, and thus providing a substantial reduction in the computational complexity. %%%%% for practical implementations.

%%%%\subsubsection{Signalling Analysis}
%%%%We consider the C-RAN architecture where a centralized BBU is aware of the instantaneous arrivals, queue backlogs, and channel-states of all users, for the design of optimal beamforming vectors~$\mathbf{f}_{b,k}(t)$ for all $b\in\mathcal{B}$, $k\in\mathcal{K}$. Hence, in the proposed methods, there is no additional signalling exchange and cooperating overhead among the CoMP \acp{RRU}. Therefore, proposed algorithms provide a low-complexity solution for the practical implementations, and these are efficiently supported by the C-RAN architecture in 5G systems.  

%% --------------------------------------  Dynamic Subset Selection  -------------------------------------- %%
\vspace*{-3.5px}
\section{Dynamic Serving Subset Selection}
\label{subsec:DynSubSet_Selection}
 \vspace*{-2.5px}

We assume that the blockers are randomly distributed and independent for each time slot. Further, the position of each blocker and/or blockage events can not be known during the downlink data transmission phase. Therefore, to improve system reliability under these uncertainties of \ac{mmWave} radio channel, we preemptively underestimate the achievable rate of each user, assuming that a portion of \ac{CoMP} links would be blocked during the data transmission phase~(see Section~\ref{subsec:Blockage}). Let \ac{BBU} assume that each user~$k$ have at least $L_k(t)$ available links (i.e., unblocked \acp{RRU}). Then we allow BBU to proactively model the pessimistic estimate of \ac{SINR} over all possible subset combinations, e.g., by excluding the potentially blocked links, and allocate the rate to users such that transmission reliability is improved (for more details see Section~\ref{subsec:Blockage}). Hence, the user-specific \ac{CoMP} subset combinations $L_k(t) \in [1, \ |\mathcal{B}_k|] \ \forall k$ is a design parameter, which can be tuned for each time slot~$t$, e.g., based on available queue backlogs and channel  information, to achieve desired rate, reliability, and user-specific latency requirements.

Let $\varrho_{k}(t) \in [0,1]$ denote the blockage probability of $k$th \ac{UE} during time slot~$t$. Then, for a fixed subset combinations~$L_k(t)$, the success probability 
%%%%\footnote{The formulation can easily be extended by assuming link specific blockage~$q_{b,k}(t) \ \forall (b,k)$. Then, success probability of $k$th \ac{UE} can be computed as $ p_k(L_k(t))\! = \! \sum\limits_{c=1}^{C(L_k(t))} \biggl(   \prod\limits_{b\in\mathcal{B}_k\comb} \! \big(1-q_{b,k}(t)\big)\! \times \! \prod\limits_{b\in\mathcal{D}_k\comb} \! q_{b,k}(t) \biggr) \ \ \forall k\in\mathcal{K}$~\cite{Dileep_Globecom2019}.} 
of $k$th \ac{UE} can be approximated as (we refer the reader to \cite[Section III]{Dileep_TWC2021} for details)
\begin{equation}
    \label{eq:SuccessProb}
    p_k\bigl(L_k(t)\bigr) \! = \!\!\!\! { \sum\limits_{l=0}^{|\mathcal{B}_k|-L_k(t)}}  \binom{|\mathcal{B}_k|}{l} \bigl(1-\varrho_{k}(t)\bigr)^{|\mathcal{B}_k|-l}  \bigl(\varrho_{k}(t)\bigr)^l.
\end{equation}
Since all users are independent, the outage probability of $k$th \ac{UE} can be expressed as
\begin{equation}
\label{eq:Pout_Formula}
	\mathrm{P}^{\mathrm{out}}_k\bigl(L_k(t)\bigr) = 1 - p_k\bigl(L_k(t)\bigr),  \ \ \forall k \in \mathcal{K}.
\end{equation}
%

%%%%From expression~\eqref{eq:Pout_Formula}, we can observe the impact of parameter~$L_k(t)$ on system reliability~\eqref{eq:P1-C1}, user-specific latency~\eqref{eq:P1-C2} and achievable rates~\eqref{eq:rate}. 
%
From expression~\eqref{eq:Pout_Formula}, we can observe that the outage $\mathrm{P}^{\mathrm{out}}_k\bigl(L_k(t)\bigr)$ is a monotonically increasing function of parameter~$L_k(t)$.
As an example, we can improve the system reliability and avoid the outage 
%%%%by using smaller values of $L_k(t)$, i.e., 
by preemptively assuming that a significant portion of the available \ac{CoMP} \acp{RRU} (i.e., $|\mathcal{B}_k|-L_k(t), \ \forall k$) are potentially blocked. However, this pessimistic assumption on available links may lead to a lower \ac{SINR} estimate (see Section~\ref{subsec:Blockage}), and hence, a lower rate to each user~\eqref{eq:rate}. Thus, to ensure the user-specific latency requirements~\eqref{eq:P1-C2}, the network consumes more power, and attempt to increase the instantaneous rate over the available subset of \acp{RRU}, $\mathcal{B}\comb_k \ \forall k \in \mathcal{K}$.

Conversely, a less pessimistic assumption on subset size (i.e., a higher value of $L_k(t) \ \forall k$) can provide higher instantaneous \ac{SINR}~\eqref{eq:SINR-Comb}, but it can be more susceptible to the outage, and results in less stable connectivity. Moreover, these outage events will eventually increase the queue-backlogs~\eqref{eq:Queue_Dynamics}, i.e., due to unsuccessful downlink data transmissions. Thus, to guarantee the desired average latency requirements~\eqref{eq:P1-C2}, the network consumes more power, and tries to increase the achievable rates in the following time slots. Clearly, there is a trade-off between reliable connectivity and sum-power performance, while ensuring the latency requirements.

Therefore, for each time slot~$t$, first we need to choose parameter $L_k(t)\in[1, \ |\mathcal{B}_k|]$, and then solve problem~\eqref{eq:P1-prob-2} over a given subset combinations~$C(L_k(t))$ for all $k\in\mathcal{K}$. 
Specifically, the parameter~$L_k(t)$, such that the solution of problem~\eqref{eq:P1-prob-2} satisfy~\eqref{eq:P1-C2} with the minimum sum-power is of our interest.  %%%%, e.g., the average queue backlogs of $k$th user is less than $Q^{\mathrm{th}}_k$ with probability $1-\epsilon$.  
Thus, we can observe that the constraint~\eqref{eq:P1-C2} can be met with minimum sum-power, if the success probability of each user~$k$ satisfy $p_k\bigl(L_k(t)\bigr) \geq 1-\epsilon, \ \forall k\in\mathcal{K}$.
Hence, the parameter $L_k(t)$ for $k$th \ac{UE} during time slot~$t$ can be computed by solving following: 
%
%%\begin{align}
%%    \label{eq:Lk_Compute}
%%    \underset{L_k(t)}{\mathrm{min}} \ \ p_k\big(L_k(t)\big) \geq 1 - \epsilon, \ \ \forall L_k(t) \in [1 , \       |\mathcal{B}_k|], \ \forall k\in\mathcal{K}.
%%\end{align}
%
%
\begin{subequations}
\label{eq:Lk_Compute}
\begin{align}
	\displaystyle 
	\begin{split}
	 %%\hspace{-4mm} \underset{L_k(t)}{\mathrm{min}} \ \ & \Bigl( p_k\big(L_k(t)\big) - (1 - \epsilon) \Bigr)
	 \underset{L_k(t)}{\mathrm{min}} \ \ &  p_k\big(L_k(t)\big) 
	 \end{split} \\
	 \mathrm{s. t.} \ \
	\begin{split}
    & \displaystyle p_k\big(L_k(t)\big) \geq 1 - \epsilon, \ \ \forall L_k(t) \in \bigl[1 , \ |\mathcal{B}_k| \bigr].
    \end{split}
	\label{eq:P1-C1-Lx}
\end{align}
\end{subequations}

It should be noted that the blockage probability $\varrho_{k}(t) \ \forall k$ is still an unknown parameter, and thus hinders solving \eqref{eq:Lk_Compute}. However, an approximation of blockage $\widetilde{\varrho}_k(t)\ \forall k$ can be obtained by, e.g.,  exploiting the available queue backlog information at the BBU. 
For example, in the considered C-RAN architecture, the centralized BBU is aware of the instantaneous arrivals,  current queue backlogs, and channel  information, for the design of beamforming vectors. 
Thus, user~$k$ during time slot~$t$ can be in outage, if the assigned downlink rates~$r_k(t)\neq0$ and the queue backlogs grow as $Q_k(t+1) = [Q_k(t)+A_k(t)]^{+}, \ \forall k \in \mathcal{K}$ (see expression~\eqref{eq:Queue_Dynamics} and \eqref{eq:Queue_Dynamics_Blockage} for details). 
%%%%Alternatively, the outage probability can be (accurately) computed based on \ac{UE} acknowledgments. However, in the presence of random link blockages, the  mmWave feedback links are inherently unreliable and, hence, results in overestimation and increased delays~\cite{Harish-ARQ}. In fact, this is an interesting topic for future extensions.
%
Therefore, the outage of $k$th \ac{UE} during time slot~$t$ can be approximated as
\begin{eqnarray*}
\mathbbm{1}_{\mathcal{P}_k}(t) \! \! \! \! & = &  \! \! \! \left\{
    \begin{array}{ll}
        {1} & \text{ if } \ \  \bigl\{ r_k(t)\neq 0  \bigr\} \bigcap \\  & \qquad \ \ \bigl\{ Q_k(t+1) = [Q_k(t)+A_k(t)]^{+} \bigr\} , \\
        {0} & \text{ otherwise. }
    \end{array}
\right.
\end{eqnarray*}
where $\mathbbm{1}_{\mathcal{P}_k}(t)$ is indicator function.  
Alternatively, the outage event can be (accurately) computed based on \ac{UE} acknowledgments. However, in the presence of random link blockages, the mmWave feedback links are inherently unreliable and, hence, results in overestimation and increased delays~\cite{Harish-ARQ, Zheng_IoTJ_2019}. In fact, this is an interesting topic for future extensions.
Thus, at each time slot~$t$, BBU exploits the available channel and queue backlogs information of each user~$k$ to compute the approximated blockage as
%
%%%%\vspace{-1.5pt}
\begin{equation}
\label{eq:Prob_Approximate}
    %\widetilde{\varrho}_k(t) = \frac{1}{\delta_k}\sum_{i=1}^{\delta_k}
    \widetilde{\varrho}_k(t) = \frac{1}{\delta_k}\sum_{i=t-\delta_k}^{t-1} \mathbbm{1}_{\mathcal{P}_k}(i), %%%\  \ \kappa_k\in\bigl\{1,2, \ldots, (t-\delta_k+1)\bigr\},
\end{equation}
where $\delta_k$ %%%% 1\!\leq\! \delta_k\! \leq (t-1), \ \forall k$ 
is maximum averaging length. Hence, using the estimated time-averaged blocking, the BBU first computes the adequate size of the subset combinations from~\eqref{eq:Lk_Compute}, and then solves problem~\eqref{eq:P1-prob-2} to obtain the  beamforming vectors.

 \vspace*{-2.5px}
\section{Simulation Results}
 \vspace*{-2.5px}
\label{sec:Sim-Result}

This section provides numerical examples to quantify the performance advantage of the proposed algorithms.
%
%
%%%%In particular, we analyze the impact of serving size $L_k$ for all $k$ on average sum-power, rates, achievable latency, and reliable \ac{mmWave} connectivity, as well as evaluated the trade-off between these performance metrics. %%%%% We further elaborate on the convergence and the performance gap between the proposed algorithms.
%
%%%%\vspace{-2px}
%%\subsection{Simulation Setup}
%%\label{subsec:Sim-Setup}
%
We consider a \ac{mmWave} based donwlink transmission with \acp{UE}~$K\!=\!4$, \acp{RRU} $B\!=\!4$, and each \ac{RRU} is equipped with a \ac{ULA} of $N\!=\!16$ antennas. Further, \acp{RRU} are placed in a $50\!\times\!50$~meters square layout (resembling, e.g., a factory-type IIoT setup), and are connected to a common \ac{BBU} in the edge cloud.  %%%%, such that all \acp{RRU} coherently serve each \ac{UE}.       
%
%%%%It should be noted, in the considered C-RAN architecture, \ac{RRU} performs only limited radio operations while all the digital signal processing functionalities are implemented in the common \ac{BBU}~\cite{CRAN-2015}. Therefore, such centralized baseband processing provides better synchronization and inference coordination among distributed \acp{RRU}, and thus enables efficient implementation for \ac{JT}-\ac{CoMP} scenarios~\cite{CRAN-2015}.
%
All~single antenna \acp{UE} are randomly dropped within the square, thus each~\ac{UE} has a different path-gain and angle with the \acp{RRU}.

%% ----- %%%% ----- %%%% ----- %%%% ----- %%

The \ac{mmWave} channel $\mathbf{h}_{b,k}(t)$ between a \ac{RRU}-\ac{UE} pair $(b,k)$ is based on sparse geometric model~\cite{Sayeed-SpatialChannel-2002}, and defined as 
\begin{equation}
    \label{eq:channel-def}
\mathbf{h}_{b,k}(t) = \sqrt{\frac{N}{M}} \sum_{m=1}^{M} {\omega}_{b,k}(t) d_{b,k}^{-\psi_m(t)}(t) \mathbf{a}_T\herm(\phi_{b,k}^m(t)), 
\end{equation}
where $M$ is the number of independent paths, which we set as $M=3$, and  $\omega_{b,k}(t)$ is random complex gain with zero mean and unit variance. The distance between \ac{RRU}-\ac{UE} pair is represented with $d_{b,k}$, and the notation $\psi_m$ denote a random path-loss exponent. In the simulation, we consider $\psi_m(t) \! \in \! [2, \ 6], \ \forall (m,t)$.   %%for all~$m=1.\ldots,M$.  
The array response vector for {ULA} is represented with $\mathbf{a}_T(\phi_{b,k}^m(t)) \in \mathbb{C}^{N \times 1}$, and relative to the boresight of the \ac{RRU} antenna array, the \ac{AoD} for each path is uniformly distributed, i.e., $\phi_{b,k}^m(t)\in[-\pi/2, \ \pi/2], \ \forall (m, t)$.
For simplicity, we assume a probabilistic blockage model~\cite{Bai-Blockage-2014, Renzo-2015-BinaryChannel}, where the radio channel between RRU-UE pair during downlink transmission phase, is either fully available, i.e., as in~\eqref{eq:channel-def} or completely blocked, i.e.,~$\{\mathbf{h}_{b,k}(t) = \mathbf{0}\}$, with the probability of $q \in [0,1]$ for all $ b\in\mathcal{B}$ and $ k\in\mathcal{K}$.

Recall that to improve the communication reliability and avoid outage under these uncertainties of \ac{mmWave} radio channel, we use parameter $L_k(\leq |\mathcal{B}_k|)$ in problem~\eqref{eq:P1-prob}, and proactively model the \ac{SINR} over the link blockage combinations (see Section~\ref{subsec:Blockage}). For simplicity, but without loss of generality, we assume identical parameters for each user~$k$, i.e., serving set $|\mathcal{B}_k|\! =\! 4$, $L_k\! =\! L$, arrival $A_k\sim\mathrm{Pois}(\lambda)$ with $\lambda=3.5$ bits/slot and maximum queue length $Q_k^{\mathrm{th}}=5$ bits with tolerable probability $\epsilon\!=\!0.1$ in problem~\eqref{eq:P1-prob}~\cite{Dileep_Globecom2020}. In the simulations, we set the frequency $f_c\!=\!28$ GHz, the step size  $\beta_e\!=\!0.01$ in expression~\eqref{eq: KKT-i_FP, ek}. %%%%%, and $\tau\!=\!0.6$ in~\eqref{eq: KKT-i_FP, nu_i}.

The outage event occurs if the instantaneous transmit rate~$r_k(t)$ exceeds the supported rate\footnote{For a give time slot $t$, let $\mathcal{S}_k(t)=\{ \gamma_k^{\star}(t), \mathbf{f}_k^{\star}(t) \}_{k\in\mathcal{K}}$ denote solution of problem~\eqref{eq:P1-prob}. Then, for each user~$k$, the transmission rate is given by $r_k(t)=\log_2(1\!+\!\gamma_k^{\star}(t))$. However, the actual supported rate (i.e., link capacity) for $k$th user depends on the obtained beamformers $\{\mathbf{f}_k^{\star}(t) \}_{\forall k\in\mathcal{K}}$ and current channel state $\{\mathbf{h}_{b,k}(t)\}_{b\in \mathcal{B}, k \in\mathcal{K}}$. However, channel  can not be exactly known to the \ac{BBU} during data transmission phase due to random blockages. Thus, the supported rate can be calculated using the actual \ac{SINR} values~\eqref{eq:SINR}, i.e., $c_k(t)\!=\!\log_2( 1\!+\!\Gamma_k(\mathbf{F}^{\star}(t)) ), \ \forall k$, and these rates are unknown to the \ac{BBU}.} $c_k(t)$ for all $ k\in\mathcal{K}$. 
Then, the queue dynamics  $Q_k(t)$ in~\eqref{eq:Queue_Dynamics} can be expressed~as
\begin{equation}
\label{eq:Queue_Dynamics_Blockage}
    Q_k(t+1) \! = \! \bigl[Q_k(t) - r_k(t)\mathbbm{1}_{\scriptsize \{r_k(t) \leq c_k(t)\}} + A_k(t)\bigr]^{+}, \ \forall k,
\end{equation}
where $\mathbbm{1}_{\{ \cdot \}}$ is an indicator function. Specifically, expression~\eqref{eq:Queue_Dynamics_Blockage} implies that queue backlogs also increase with each unsuccessful downlink transmission due to random blockages. %%%%Recall that to improve communication reliability, we use subset size~$L$ for the pessimistic estimate of achievable SINR. 
In our study, we will make use of subset size~$L$ to analyze the network performance. 
For the baseline methods, we consider \ac{CB} (i.e., $|\mathcal{B}_k| = 1, \ \forall k$)~\cite{Antti-OntheValue-2009} and full-JT (i.e., $L_k = |\mathcal{B}_k|, \ \forall k$)~\cite{Kaleva-DecentralizeJP-2018} based downlink beamformer designs.

%% ----- %%%% ----- %%%% ----- %%%% ----- %%

\subsection{Convergence Analysis}
\label{subsec:SimConvergance}

\begin{figure}[t]
 \centering
 \setlength\abovecaptionskip{-0.25\baselineskip}
 \includegraphics[trim=0.25cm 0.08cm 0.25cm 0.2cm, clip, width=1\linewidth]{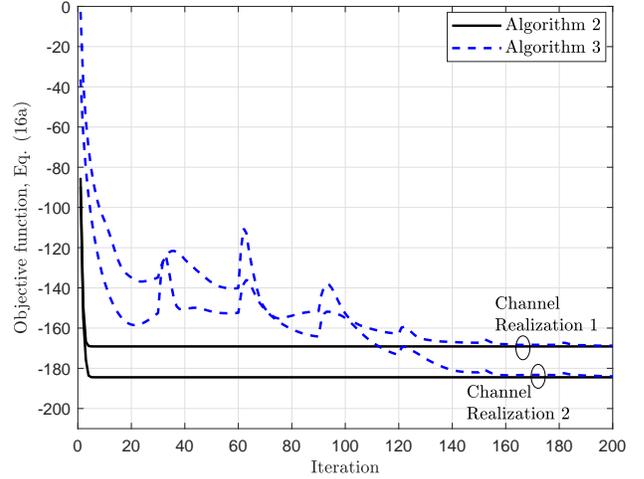}
 \caption{Convergence performance of Algorithm~\ref{algFP} and Algorithm~\ref{algKKT_FP}. }
 \label{fig:Convergance}
%%  \vspace*{-10pt}
 \end{figure}

In Fig.~\ref{fig:Convergance}, we examine the convergence behavior of proposed  iterative algorithms for given randomly generated channel realizations. For simplicity but without loss of generality, we set parameters, $V=1$, $L=3$, and $q=0$. Note that the solution of Algorithms~\ref{algFP} is obtained directly by using the convex optimization toolbox,~$\mathrm{SeDuMi}$~\cite{cvx}. In contrast, Algorithm~\ref{algKKT_FP} is solved from a system of closed-form \ac{KKT} optimality conditions in an iterative manner (see~\eqref{eq:KKT-i_FP iteration}). We can observe that both algorithms converge to the same local solution, on average, within a fairly small number of iterations. It is worth highlighting, in general, the convergence of Algorithm~\ref{algKKT_FP}  cannot be guaranteed to be monotonic due to subgradient updates~\eqref{eq: KKT-i_FP, ek}, and it can be tuned in accordance with the system parameters. We refer the reader to~\cite{boyd2003subgradient} on the convergence properties of the subgradient approach with different step size rules. The Algorithm~\ref{algFP} provides monotonic and faster convergence in terms of required approximation point updates. On the contrary, Algorithm~\ref{algKKT_FP} achieves comparable
performance with a significant reduction in the per-iteration computational complexity, which can be useful for, e.g., hardware constrained IoT devices with limited processing capabilities in IIoT and factory automation scenarios.

%

%Next,  we compare the average sum-power performance of both algorithms with increasing transmit antennas. %%%%the low-complexity KKT based iterative method 
%%%%based on iteratively solving a system of closed-form \ac{KKT} optimality conditions 
%%%with the solution obtained directly by the convex optimization toolbox~\cite{cvx}. 
%
%It can be concluded from Fig.~\ref{fig:SumPowerWithN} that both algorithms achieve the similar performance, and the resulting gap is due to insufficient convergence because of a fixed number of maximum iterations (i.e., $\mathrm{i}\!=\!200$ approximation point updates). 
%
%Furthermore, the converged values of different algorithms may differ depending on the starting point. %%%%, as only stationary point convergence can be guaranteed in all cases.. 
%
%Therefore, the proposed \ac{KKT} based iterative methods provide a low-complexity implementation, without any significant degradation in the system performance. In the following, the numerical examples are provided with Algorithm~\ref{algFP} due to space limitation, but we can observe similar behavior for both algorithms, as also shown in Fig.~\ref{fig:SumPowerWithN}. 

\subsection{Impact of the trade-off parameter~$V$}
\label{subsec:SimWithV}

%%%%%%%%%%%%%%%%%%%%%%%%%%%%%%%%%%%%%%%%%%%%%%%%%%%%%%%%%%%%%%%%%%%%%%%%%%%%%%%%%
\begin{comment}
%% ----- %%%% ----- %%
%
\begin{figure}[t]
\setlength\abovecaptionskip{-0.2\baselineskip}
\centering
\includegraphics[trim=0.25cm 0.15cm 0.25cm 0.2cm, clip, width=0.49\linewidth]{figsPaper/Queue_V1.eps} 
\caption{Queue backlog with $V=1$ and blockage probability of $20\%$.}
\label{fig:QueueLength-Results}
\end{figure}

\begin{figure}[t]
\setlength\abovecaptionskip{-0.2\baselineskip}
\centering
\includegraphics[trim=0.25cm 0.15cm 0.25cm 0.2cm, clip, width=0.49\linewidth]{figsPaper/SumPowerVSV.eps} 
\caption{Sum-power with increasing $V$ and blockage probability of $20\%$.}
\label{fig:Power-Results-V}
\end{figure}
%% ----- %%%% ----- %%
\end{comment}
%%%%%%%%%%%%%%%%%%%%%%%%%%%%%%%%%%%%%%%%%%%%%%%%%%%%%%%%%%%%%%%%%%%%%%%%%%%%%%%%%

%
%
%%\begin{figure}[t]
%%\begin{minipage}[c]{0.49\textwidth}
 
 \begin{figure}[t]
 \centering
 \setlength\abovecaptionskip{-0.25\baselineskip}
 \includegraphics[trim=0.25cm 0.08cm 0.25cm 0.2cm, clip, width=1\linewidth]{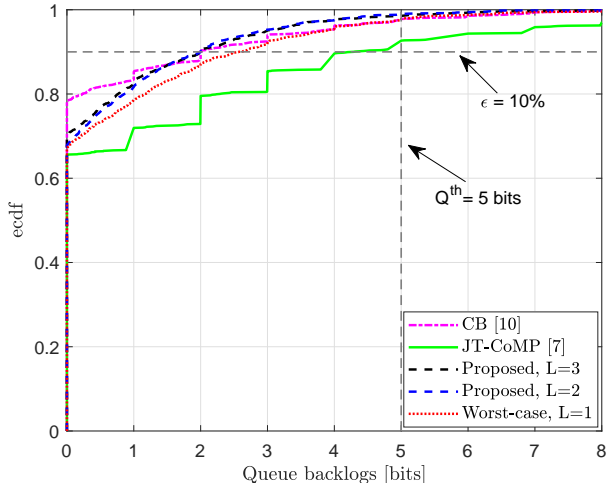}
 \caption{Queue backlogs with $V=1$ and blockage of $10\%$.}
 \label{fig:QueueLength-Results}
%%   \vspace*{-10pt}
\end{figure}

%%\end{minipage}
%%\hspace{1mm}
%%\begin{minipage}[c]{0.49\textwidth}

\begin{figure}[t]
 \centering
 \setlength\abovecaptionskip{-0.25\baselineskip}
 \includegraphics[trim=0.25cm 0.08cm 0.25cm 0.2cm, clip, width=1\linewidth]{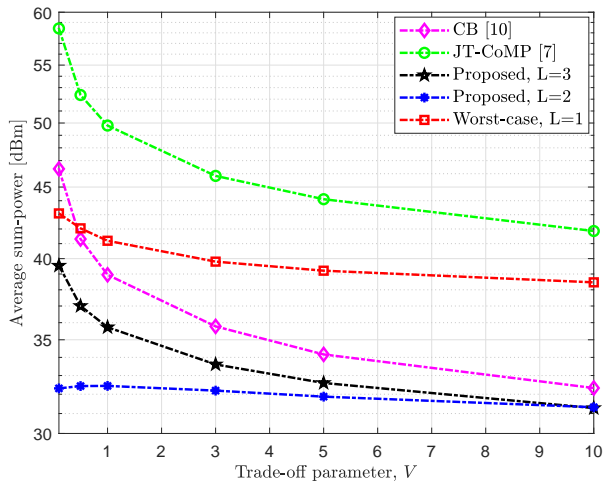}
 \caption{Average sum-power with increasing $V$ and blockage of $10\%$.}
 \label{fig:Power-Results-V}
%%   \vspace*{-10pt}
\end{figure}

%%\end{minipage} 
%%\end{figure}
%
%

First, in Fig.~\ref{fig:QueueLength-Results}, we illustrate the  latency performance with trade-off parameter $V\!=\!1$ and blockage $q\!=\!10\%$. The result shows that our proposed and baseline methods satisfy the maximum queue backlogs of each user~$k$ (i.e., $Q_k^{\mathrm{th}}\!=\!5$) within the allowable queue tolerance level $\epsilon\!=\!0.1$. Thus, problem~\eqref{eq:P1-prob} is feasible, and the proposed convex relaxations still allow to achieve the desired user-specific latency requirements (i.e., constraint~\eqref{eq:P1-C2} is  met). 
However, our proposed  beamformer designs, i.e., by considering a pessimistic estimate of rates over the subset combinations of potentially blocked \ac{CoMP} \acp{RRU}, substantially improve the average sum-power performance, while ensuring the same latency requirements, as shown in Fig.~\ref{fig:Power-Results-V}. As an example, for the parameter $V\!=\!1$ and  $L\!=\!2$, our proposed method improves the average sum-power performance by $8$~dBm and $18$~dBm compared to baseline \ac{CB} and full-{JT}, respectively. 
Hence, the proposed methods significantly outperform the conventional full-\ac{JT}~\cite{Kaleva-DecentralizeJP-2018} and \ac{CB}~\cite{Antti-OntheValue-2009} based downlink beamformer design, and thus provides power-efficient and low-latency \ac{mmWave} communication.

%
%
%%\begin{figure}[b]
%%\begin{minipage}[c]{0.49\textwidth}

\begin{figure}[t]
 \centering
 \setlength\abovecaptionskip{-0.25\baselineskip}
 \includegraphics[trim=0.25cm 0.08cm 0.25cm 0.2cm, clip, width=1\linewidth]{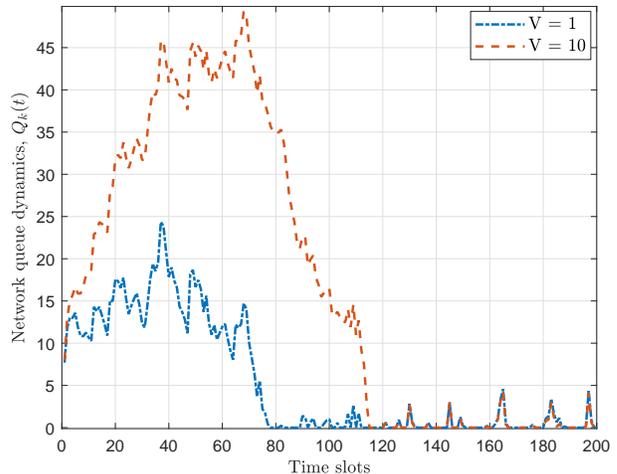}
 \caption{Network queue dynamics with blockage of $10\%$.}
 \label{fig:Queue-Qk}
%%   \vspace*{-10pt}
\end{figure} 
 
%%\end{minipage}
%%\hspace{1mm}
%%\begin{minipage}[c]{0.49\textwidth}

\begin{figure}[t]
 \centering
 \setlength\abovecaptionskip{-0.25\baselineskip}
 \includegraphics[trim=0.25cm 0.08cm 0.25cm 0.2cm, clip, width=1\linewidth]{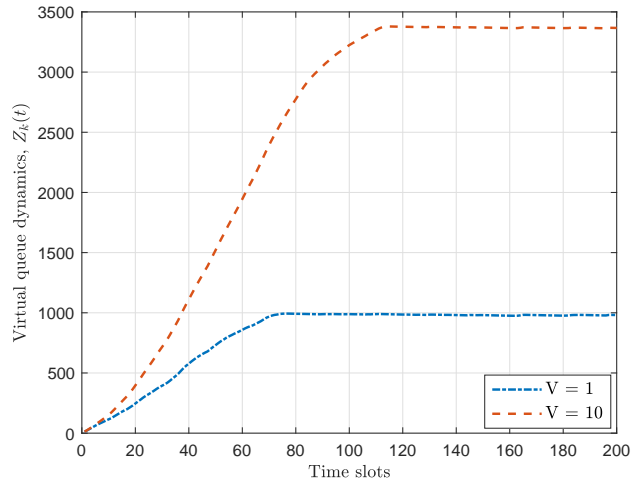}
 \caption{Virtual queue dynamics with blockage of $10\%$.}
 \label{fig:Queue-Zk}
%%   \vspace*{-10pt}
\end{figure}

%%\end{minipage} 
%%\end{figure}
%
%

Further, it can be observed from Fig.~\ref{fig:Power-Results-V} that for a fixed queue length constraint, the average sum-power decreases with the increase in the value of trade-off parameter~$V$. This behavior is expected, since higher values of parameter~$V$ linearly emphasize the minimization of the sum-power objective over the queue length, until the queue backlogs become substantially larger than the sum-power objective values (see~\eqref{eq:P1-2}). %%%%% Thus, problem~\eqref{eq:P1-prob-2} may become infeasible for large values of trade-off parameter~$V$.

Fig.~\ref{fig:Queue-Qk} and Fig.~\ref{fig:Queue-Zk} shows the evolution of the network queues~$\{Q_k(t)\}$ and the associated virtual queues~$\{Z_k(t)\}$ over time with the blockage~$q=10\%$ and the parameter $L=3$, for user~$k=1$. Note that, we can observe similar behavior for all other users, but these are not included due to space limitations. %%%for the trade-off parameter $V=3$ and $V=10$. 
It can be concluded from Fig.~\ref{fig:Queue-Qk} that the queue backlogs increases differently for different values of trade-off parameters~$V$, until it saturate and reaches a certain value (e.g., for $V=10$ around $t=115$ time slots), and then it oscillates, so as the constraint $\mathrm{Pr}\bigl\{ Q_k(t) \geq Q_k^{\mathrm{th}} \bigr\} \! \leq \! \epsilon$, is ensured, i.e., to achieve the average user-specific latency requirements. This is mainly because of the negative drift property of the Lyapunov function~\cite[Ch. 4.4]{neely2010stochastic}. Thus, the stability of associated virtual queues~$\{Z_k(t)\}$, i.e., as in Fig.~\ref{fig:Queue-Zk}, ensures that the network queues are bounded, and achieves the desired queue backlogs performance. %%%%, i.e., constraint~\eqref{eq:P1-C2} is satisfied. 

%% ----- %%%% ----- %%%% ----- %%%% ----- %%

\subsection{Dynamic selection of parameter $L$}
\label{subsec:SimWithBlockage}

%
%
%%\begin{figure}[b]
%%\begin{minipage}[c]{0.49\textwidth}

\begin{figure}[t]
 \centering
 \setlength\abovecaptionskip{-0.25\baselineskip}
 \includegraphics[trim=0.25cm 0.08cm 0.25cm 0.2cm, clip, width=1\linewidth]{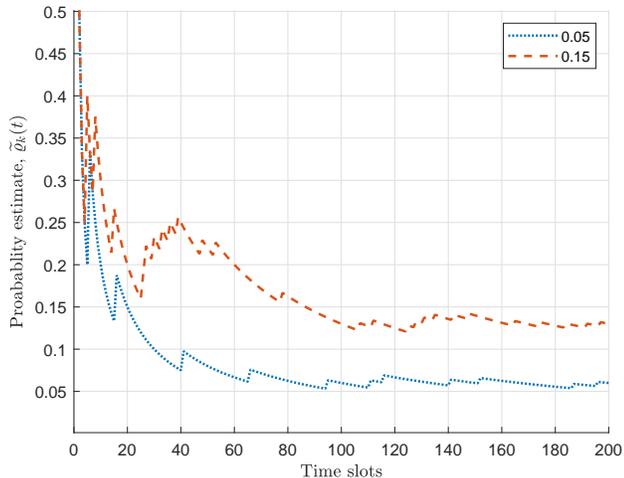}
 \caption{Approximation of blockage probability using~\eqref{eq:Prob_Approximate} .}
 \label{fig:Blockage-Esti}
%%   \vspace*{-10pt}
\end{figure}

%%\end{minipage}
%%\hspace{1mm}
%%\begin{minipage}[c]{0.49\textwidth}

\begin{figure}[t]
 \centering
 \setlength\abovecaptionskip{-0.25\baselineskip}
 \includegraphics[trim=0.25cm 0.08cm 0.25cm 0.2cm, clip, width=1\linewidth]{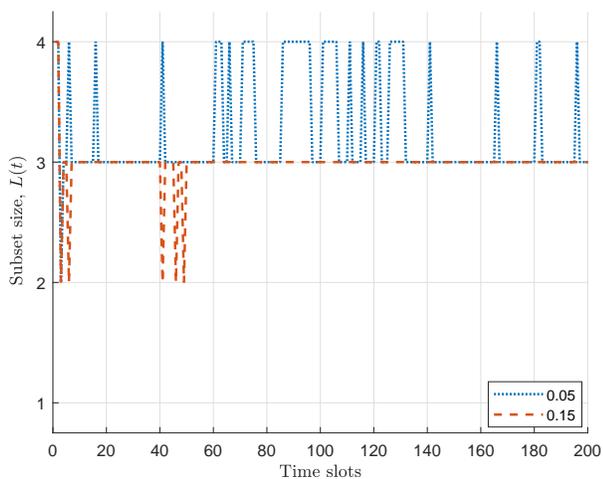}
 \caption{Dynamic selection of subset size $L(t)$ by solving~\eqref{eq:Lk_Compute}.}
 \label{fig:L-Variation}
%%   \vspace*{-10pt}
\end{figure}

%%\end{minipage} 
%%\end{figure}
%
%

%%%%%%%% Recall that to adopt with uncertainties of \ac{mmWave} radio channel, we preemptively model the lower-bound \ac{SINR} over the link blockage combinations (see Section~\ref{subsec:Blockage}). Thus, the serving set size $L(t)$ is a design parameter, which can be tuned for each time slot~$t$, e.g., based on available queue backlogs, instantaneous arrival, and channel state information, in order to achieve desired rate, reliability, and sum-power performance. 

Next, we investigate the dynamic selection of the serving set size, i.e., parameter~$L(t)$ for each time slot~$t$, which is obtained by solving~\eqref{eq:Lk_Compute}. To do that, first   
%
%%%%fig.~\ref{fig:Blockage-Esti} shows that 
%%%% the time-averaged blockage computed from expression~\eqref{eq:Prob_Approximate} with %%$\kappa_k=1$ and
%%%% $\delta_k=(t-1)$ is shown in Fig.~\ref{fig:Blockage-Esti}. 
%
in Fig.~\ref{fig:Blockage-Esti}, we show the time-averaged blockage computed from expression~\eqref{eq:Prob_Approximate} with the maximum averaging length $\delta_k=\mathrm{min}\{\tau, \ (t-1)\}$, and we set $\tau=50$. 
It can be concluded that the estimate of the blockage in~\eqref{eq:Prob_Approximate} closely matches with actual blockage probability, within a moderately small number of random channel realizations. Thus, expression~\eqref{eq:Prob_Approximate} provides a fair approximation, and the resulting gap is mainly due to unpredictable random blockage events and uncertainties in the \ac{mmWave} radio channel.

%%%%%Next, in Fig.~\ref{fig:L-Variation}, we investigate the dynamic selection of subset size ${L_k(t)}$, which is obtained by solving~\eqref{eq:Lk_Compute}. 

Further, it can be observed from Fig.~\ref{fig:L-Variation} that the instantaneous choice of serving set size, i.e., to meet the success probability~\eqref{eq:P1-C1-Lx}, mainly depends on the  accuracy of estimated blockage. Furthermore, the oscillations in Fig.~\ref{fig:L-Variation} is due to limited (and discrete) choices of parameter $L_k(t) \in [1, \ |\mathcal{B}_k|], \ \forall k$, and possibly the minimum subset size satisfying~\eqref{eq:P1-C1-Lx} can be in between these integer values. It is worth highlighting that the problem~\eqref{eq:Lk_Compute} aims at finding serving set size, such that solving problem~\eqref{eq:P1-prob-2} satisfy the~\eqref{eq:P1-C2} with minimum sum-power. Therefore, large values of $L$ may result in lower sum-power, but it 
%%%%can be more venerable to the 
may lead to higher 
outage, and thus it dynamically switches to a more pessimistic \ac{SINR} estimate, i.e., to lower values of $L$, in the following time slot, to meet the average latency requirements. Hence, there is a trade-off between average sum-power, achievable rate, and reliability, as will be demonstrated in the following.

%%%%%%%%%%%%%%%%%%%%%%%%%%%%%%%%%%%%%%%%%%%%%%%%%%%%%%%%%%%%%%%%%%%%%%%%%%%%%%
\begin{comment}
%% ----- %%%% ----- %%

\begin{figure}[t]
\setlength\abovecaptionskip{-0.2\baselineskip}
\centering
\includegraphics[trim=0.25cm 0.15cm 0.25cm 0.2cm, clip, width=0.49\linewidth]{figsPaper/SumPower_Plot.eps} 
\caption{Sum-power with increasing $V$ and blockage probability of $20\%$.}
\label{fig:Power-Results-Blockage}
\end{figure}

\begin{figure}[t]
\setlength\abovecaptionskip{-0.2\baselineskip}
\centering
\includegraphics[trim=0.25cm 0.15cm 0.25cm 0.2cm, clip, width=0.49\linewidth]{figs/ecdf_Log2PlotFinal.eps} 
\caption{Effective user-rate with $V=1$, blockage probability $q_{b,k}=10\%$ (solid~line) and blockage probability $q_{b,k}=30\%$ (dotted line).}
\label{fig:Rate-Results}
\end{figure}

%% ----- %%%% ----- %%
\end{comment}
%%%%%%%%%%%%%%%%%%%%%%%%%%%%%%%%%%%%%%%%%%%%%%%%%%%%%%%%%%%%%%%%%%%%%%%%%%%%%%

%
%
%%\begin{figure}[b]
%%\begin{minipage}[c]{0.49\textwidth}

\begin{figure}[t]
 \centering
 \setlength\abovecaptionskip{-0.25\baselineskip}
 \includegraphics[trim=0.25cm 0.08cm 0.25cm 0.2cm, clip, width=1\linewidth]{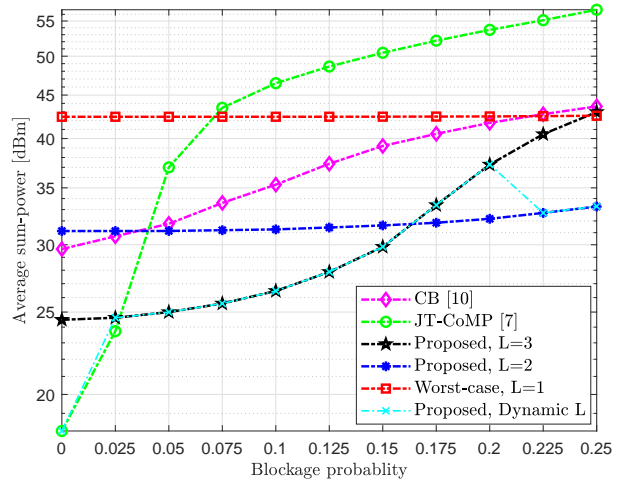}
 \caption{Average sum-power with increasing blockage and $V\!=\!1$.}
 \label{fig:Power-Results-Blockage}
%%   \vspace*{-10pt}
\end{figure}

%%\end{minipage}
%%\hspace{1mm}
%%\begin{minipage}[c]{0.49\textwidth}

\begin{figure}[t]
 \centering
 \setlength\abovecaptionskip{-0.25\baselineskip}
 \includegraphics[trim=0.25cm 0.08cm 0.25cm 0.2cm, clip, width=1\linewidth]{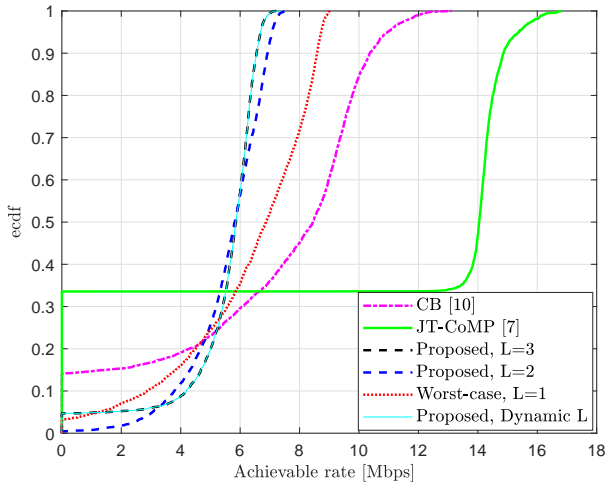}
 \caption{Achievable user-rate with $V\!=\!1$, blockage $q=10\%$.}
 \label{fig:Rate-Results}
%%   \vspace*{-10pt}
\end{figure}

%%\end{minipage} 
%%\end{figure}
%
%

%%%%%%%%%Next, we investigate the impact of different values of blockage probability on constraint~\eqref{eq:P1-C1}. average sum-power and achievable rate in Fig.~\ref{fig:Power-Results-Blockage} and Fig.~\ref{fig:Rate-Results}, respectively.  
%

Fig.~\ref{fig:Power-Results-Blockage} illustrates the impact increasing blockage probability on constraint~\eqref{eq:P1-C1} and sum-power performance. For example, the worst-case pessimistic assumption on available links (i.e., $L=1$) leads to a lower \ac{SINR} estimate. Thus, to ensure the user-specific latency requirements~\eqref{eq:P1-C2}, the network strives to increase the user's rate over the available subset of \acp{RRU} $\mathcal{B}\comb_k \ \forall k \in \mathcal{K}$, i.e., by consuming higher network power. Thus, it results in relatively lower average sum-power performance, as shown in Fig.~\ref{fig:Power-Results-Blockage}.  
Conversely, a least pessimistic assumption on subset size (i.e., \ac{JT}-\ac{CoMP}, $L=B$) can provide higher instantaneous \ac{SINR}~\eqref{eq:SINR-Comb}, but it is  more susceptible to the outage with the slight increase in blockage probability. Moreover, these outage events will eventually increase the queue-backlogs~\eqref{eq:Queue_Dynamics_Blockage}, i.e., due to unsuccessful data transmissions. Thus, to guarantee the average latency requirements~\eqref{eq:P1-C2}, the network consumes more power, and attempts to increase the achievable rates during potentially unblocked events. Thus, it results in a lower sum-power performance with increasing blockage, as shown in Fig.~\ref{fig:Power-Results-Blockage},  
%
%
%%%%%%%%%%For example, with blockage probability~$q=0.1$ and parameter~$L=3$, our proposed method improves the sum-power by $15$~dB and $8$~dB compared to full-\ac{JT} and \ac{CB}, respectively.
%
%
Furthermore, our proposed dynamic choice of serving set size ensures the same average latency requirements with minimum sum-power, and efficiently adopts with the uncertainties of \ac{mmWave} radio channel and unpredictable blockage events. Note that the resulting gap in the dynamic subset selection relative to the lower envelope is mainly due to the limited and discrete choices of parameter $L$, and the approximation error of the time-averaged blockage in~\eqref{eq:Prob_Approximate}, as also illustrated in Fig.~\ref{fig:Blockage-Esti}.

Fig.~\ref{fig:Rate-Results} illustrates the trade-offs between rates and reliable connectivity with parameter $V\!=\!1$ and blockage $q\!=\!10\%$. However, similar behavior can be observed for different parameter settings, which is not included due to space limitations.
%
%%%%%%For example, the use of smaller subset size~$L$ provides a pessimistic \ac{SINR} estimate, and hence lower rate to each user (see~\eqref{eq:rate}). However, pessimistic SINR estimate over the link blockage combinations greatly improves the outage performance. Thus, it leads to more stable and resilient connectivity in the presence of random blockages.
%
It can concluded from Fig.~\ref{fig:Rate-Results} that the outage is decreased from $35\%$ to less than $0.5\%$ by changing parameter $L$ from $4$ (full \ac{JT}-\ac{CoMP}) to $2$ (proposed), in problem~\eqref{eq:P1-prob}. Thus, the pessimistic SINR estimate over the link blockage combinations greatly improves the outage performance.
Clearly, there is a trade-off between reliable connectivity, rate, and sum-power performance. More specifically, for a given queue length (i.e., latency requirements), we can guarantee a user-rate with minimum  sum-power and vice-versa. However, compared to the baseline schemes, the proposed method provides power-efficient, high-reliability, and low-latency mmWave communication.
%

%%%%%%%%%%%%%%%%%%%%%%%%%%%%%%%%%%%%%%%%%%%%%%%%%%%%%%%%%%%%%%%%%%%
\begin{comment}
\begin{figure}[b]
\setlength\abovecaptionskip{-0.2\baselineskip}
\centering
\includegraphics[trim=0.25cm 0.15cm 0.25cm 0.2cm, clip, width=0.49\linewidth]{figsPaper/ConverganceFP_SCA.eps} 
\caption{Sum-power with increasing $V$ and blockage probability of $20\%$.}
\label{fig:Convergance}
\end{figure}
\end{comment}
%%%%%%%%%%%%%%%%%%%%%%%%%%%%%%%%%%%%%%%%%%%%%%%%%%%%%%%%%%%%%%%%%%%

%%\vspace{-2.5px}
%% CONCLUSIONS -------------------------------------------
\section{Conclusion}
%%\vspace{-2.5px}
\label{sec:conclusion}
In this paper, we have studied the trade-off between reliable downlink transmission and sum-power performance in a mmWave based dynamic systems, by exploiting the multi-antenna spatial diversity and {CoMP} connectivity. We considered a time-average sum-power minimization problem subject to maximum allowable queue length constraint. We have adapted the Lyapunov optimization framework, and derived a dynamic control algorithm for the long-term time-average stochastic problem. We proposed a robust transmit beamformer design by considering a pessimistic estimate of rates and a proactive selection of the serving set combinations of available \ac{CoMP} \acp{RRU}. Furthermore, the non-convex and coupled constraints are handled using \ac{FP} techniques. The closed-form algorithm is provided by iteratively solving a system of KKT optimality conditions, while accounting for the uncertainties of \ac{mmWave} radio channel. The numerical results manifested the robustness of the proposed beamformer design in the presence of random link blockages. Specifically, the achievable rate and sum-power performance with our proposed method outperforms the baseline scenarios while ensuring user-specific latency requirements, and results in power-efficient, highly-reliable, and low-latency {mmWave} communication.

%%%%%%%%%%%%%%%%%%%%%%%%%%%%%%%%%%%%%%%%%%%%%%%%%%%%%%%%%%%%%%%%%%%%%%%%%%%%%%%%%%%%%%%%%%%%
%%\vspace{-2.5px}
%% REFERENCES --------------------------------------------
%%\vspace{-2.5px}
\bibliographystyle{IEEEtran}
\bibliography{IEEEabrv,ref_conf_short,ref_jour_short,references}

\end{document}